# Tightened Upper Bounds on the ML Decoding Error Probability of Binary Linear Block Codes

Moshe Twitto Igal Sason Shlomo Shamai

Technion – Israel Institute of Technology
Department of Electrical Engineering
Haifa 32000, Israel
E-mails: {tmoshe@tx, sason@ee, sshlomo@ee}.technion.ac.il

June 5, 2018

**Abstract**

The performance of maximum-likelihood (ML) decoded binary linear block codes is addressed via the derivation of tightened upper bounds on their decoding error probability. The upper bounds on the block and bit error probabilities are valid for any memoryless, binary-input and output-symmetric communication channel, and their effectiveness is exemplified for various ensembles of turbo-like codes over the AWGN channel. An expurgation of the distance spectrum of binary linear block codes further tightens the resulting upper bounds.

## 1 Introduction

Since the advent of information theory, the search for efficient coding systems has motivated the introduction of efficient bounding techniques tailored to specific codes or some carefully chosen ensembles of codes. The incentive for introducing and applying such bounds has strengthened with the introduction of various families of codes defined on graphs which closely approach the channel capacity with feasible complexity (e.g., turbo codes, repeat-accumulate codes, and low-density parity-check (LDPC) codes). Clearly, the desired bounds must not be subject to the union bound limitation, since for long blocks these ensembles of turbo-like codes perform reliably at rates which are considerably above the cutoff rate ($R_0$) of the channel (recalling that union bounds for long codes are not informative at the portion of the rate region above $R_0$, where the performance of these capacity-approaching codes is most appealing). Although maximum-likelihood (ML) decoding is in general prohibitively complex for long codes, the derivation of bounds on the ML decoding error probability is of interest, providing an ultimate indication of the system performance. Further, the structure of efficient codes is usually not available, necessitating efficient bounds on performance to rely only on basic features, such as the distance spectrum and the input-output weight enumeration function (IOWEF) of the examined code (for the evaluation of the block and bit error probabilities, respectively, of a specific code or ensemble).

A basic inequality which serves for the derivation of many previously reported upper bounds is the following:

$$\Pr(\text{word error} \mid \mathbf{c}) \leq \Pr(\text{word error} \mid \mathbf{c},\ \mathbf{y} \in \mathcal{R}) + \Pr(\mathbf{y} \notin \mathcal{R}) \tag{1}$$

where $\mathbf{y}$ denotes the received vector at the output of the receiver, $\mathcal{R}$ is an arbitrary geometrical region which can be interpreted as a subset of the observation space, and $\mathbf{c}$ is an arbitrary transmitted codeword. This category includes the Berlekamp tangential bound [4] where the volume $\mathcal{R}$ is a half-space separated by a plane, the sphere bound by Herzberg and Poltyrev [11] where $\mathcal{R}$ is a hyper-sphere, Poltyrev's tangential-sphere bound [14] (TSB) where $\mathcal{R}$ is a circular cone, and Divsalar's bound [6] where $\mathcal{R}$ is a hyper-sphere with an additional degree of freedom with respect to the location of its center.

Another approach is the Gallager bounding technique which provides a conditional upper bound on the ML decoding error probability given an arbitrary transmitted (length-$N$) codeword $\mathbf{c}_m$ ($P_{\mathrm{e}|m}$). The conditional decoding error probability is upper bounded by

$$P_{\mathrm{e}|m} \leq \left( \sum_{m' \neq m} \sum_{\mathbf{y}} p_N(\mathbf{y}|\mathbf{c}_m)^{\frac{1}{\rho}} \ \psi_N^m(\mathbf{y})^{1-\frac{1}{\rho}} \ \left( \frac{p_N(\mathbf{y}|\mathbf{c}_{m'})}{p_N(\mathbf{y}|\mathbf{c}_m)} \right)^{\lambda} \right)^{\rho} \tag{2}$$

where $0 \leq \rho \leq 1$ and $\lambda \geq 0$ (see [8, 19]; in order to make the presentation self-contained, it will be introduced shortly in the next section as part of the preliminary material). Here, $\psi_N^m(\mathbf{y})$ is an arbitrary probability tilting measure (which may depend on the transmitted codeword $\mathbf{c}_m$), and $p_N(\mathbf{y}|\mathbf{c})$ designates the transition probability measure of the channel. Connections between these two seemingly different bounding techniques in (1) and (2) were demonstrated in [21], showing that many previously reported bounds (or their Chernoff versions) whose derivation originally relied on the concept shown in inequality (1) can in fact be re-produced as particular cases of the bounding technique used in (2). To this end, one simply needs to choose the suitable probability tilting measure $\psi_N^m$ which serves as the "kernel" for reproducing various previously reported bounds. The observations in [21] relied on some fundamental results which were reported by Divsalar [6].

The tangential-sphere bound (TSB) of Poltyrev often happens to be the tightest upper bound on the ML decoding error probability of block codes whose transmission takes place over a binary-input AWGN channel. However, in the random coding setting, it fails to reproduce the random coding exponent [10] while the second version of the Duman and Salehi (DS2) bound, to be reviewed in the next section, does (see [21]). The Shulman-Feder bound (SFB) can be derived as a particular case of the DS2 bound (see [21]), and it achieves the random coding error exponent. Though the SFB is informative for some structured linear block codes with good Hamming properties, it appears to be rather loose when considering sequences of linear block codes whose minimum distance grows sub-linearly with the block length, as is the case with most capacity-approaching ensembles of LDPC and turbo codes. However, the tightness of this bounding technique is significantly improved by combining the SFB with the union bound; this approach was exemplified for some structured ensembles of LDPC codes (see e.g., [13] and the proof of [18, Theorem 2.2]).

In this paper, we introduce improved upper bounds on both the bit and block error probabilities. Section 2 presents some preliminary material. In Section 3, we introduce an upper bound on the block error probability which is in general tighter than the SFB, and combine the resulting bound with the union bound. Similarly, an appropriate upper bound on the bit error probability is introduced. The effect of an expurgation of the distance spectrum on the tightness of the resulting bounds is considered in Section 4. By applying the new bounds to ensembles of turbo-like codes over the binary-input AWGN channel, we demonstrate the usefulness of the new bounds in Section 5, especially for some coding structures of high rates. We conclude our discussion in Section 6. For an extensive tutorial paper on performance bounds of linear codes, the reader is referred to [19].

# 2 Preliminaries

We introduce in this section some preliminary material which serves as a preparatory step towards the presentation of the material in the following sections.

## 2.1 The DS2 Bound

The bounding technique of Duman and Salehi [7, 8] originates from the 1965 Gallager bound. Let $\psi_N^m(\underline{y})$ designate an arbitrary probability measure (which may also depend on the transmitted codeword $\underline{x}^m$). The Gallager bound [10] can then be put in the form (see [21])

$$\begin{aligned} P_{\mathrm{e}|m} &\leq \sum_{\mathbf{y}} \psi_N^m(\mathbf{y})\, \psi_N^m(\mathbf{y})^{-1}\, p_N(\mathbf{y}|\mathbf{c}_m) \left( \sum_{m' \neq m} \left( \frac{p_N(\mathbf{y}|\mathbf{c}_{m'})}{p_N(\mathbf{y}|\mathbf{c}_m)} \right)^{\lambda} \right)^{\rho} \\ &= \sum_{\mathbf{y}} \psi_N^m(\mathbf{y}) \left( \psi_N^m(\mathbf{y})^{-\frac{1}{\rho}}\, p_N(\mathbf{y}|\mathbf{c}_m)^{\frac{1}{\rho}} \sum_{m' \neq m} \left( \frac{p_N(\mathbf{y}|\mathbf{c}_{m'})}{p_N(\mathbf{y}|\mathbf{c}_m)} \right)^{\lambda} \right)^{\rho}, \qquad \forall\, \lambda, \rho \geq 0. \end{aligned} \tag{3}$$

By invoking the Jensen inequality in (3) for $0 \leq \rho \leq 1$, the DS2 bound results

$$P_{\mathrm{e}|m} \leq \left( \sum_{m' \neq m} \sum_{\mathbf{y}} p_N(\mathbf{y}|\mathbf{c}_m)^{\frac{1}{\rho}}\, \psi_N^m(\mathbf{y})^{1-\frac{1}{\rho}} \left( \frac{p_N(\mathbf{y}|\mathbf{c}_{m'})}{p_N(\mathbf{y}|\mathbf{c}_m)} \right)^{\lambda} \right)^{\rho}, \qquad 0 \leq \rho \leq 1,\ \lambda \geq 0. \tag{4}$$

Let $G_N^m(\mathbf{y})$ be an arbitrary non-negative function of $\mathbf{y}$, and let the probability density function $\psi_N^m(\mathbf{y})$ be

$$\psi_N^m(\mathbf{y}) = \frac{G_N^m(\mathbf{y})\, p_N(\mathbf{y}|\mathbf{c}_m)}{\sum_{\mathbf{y}} G_N^m(\mathbf{y})\, p_N(\mathbf{y}|\mathbf{c}_m)} \tag{5}$$

The functions $G_N^m(\mathbf{y})$ and $\psi_N^m(\mathbf{y})$ are referred to as the un-normalized and normalized tilting measures, respectively. The substitution of (5) into (4) yields the following upper bound on the conditional ML decoding error probability

$$\begin{aligned} P_{\mathrm{e}|m} \leq\ & \left( \sum_{\mathbf{y}} G_N^m(\mathbf{y})\, p_N(\mathbf{y}|\mathbf{c}_m) \right)^{1-\rho} \\ & \cdot \left( \sum_{m' \neq m} \sum_{\mathbf{y}} p_N(\mathbf{y}|\mathbf{c}_m)\, G_N^m(\mathbf{y})^{1-\frac{1}{\rho}} \left( \frac{p_N(\mathbf{y}|\mathbf{c}_{m'})}{p_N(\mathbf{y}|\mathbf{c}_m)} \right)^{\lambda} \right)^{\rho}, \quad 0 \leq \rho \leq 1, \quad \lambda \geq 0. \end{aligned} \tag{6}$$

The upper bound (6) was also derived in [6, Eq. (62)].

For the case of memoryless channels, and for the choice of $\psi_N^m(\mathbf{y})$ as $\psi_N^m(\mathbf{y}) = \prod_{i=1}^{N} \psi^m(y_i)$ (recalling that the function $\psi_N^m$ may depend on the transmitted codeword $\mathbf{x}^m$), the upper bound (4) is relatively easily evaluated (similarly to the standard union bounds) for linear block codes. In that case, (4) is calculable in terms of the distance spectrum of the code, not requiring the fine details of the code structure. Moreover, (4) is also amenable to some generalizations, such as for the class of discrete memoryless channels with arbitrary input and output alphabets.

## 2.2 The Shulman and Feder bound

We consider here the transmission of a binary linear block code $\mathcal{C}$ where the communication takes place over a memoryless binary-input output-symmetric (MBIOS) channel. The analysis refers to the decoding error probability under soft-decision ML decoding.

The Shulman and Feder bound (SFB) [20] on the block error probability of an $(N, K)$ binary linear block code $\mathcal{C}$, transmitted over a memoryless channel is given by

$$P_{\mathrm{e}} \leq 2^{-NE_{\mathrm{r}}(R + \frac{\log \alpha(\mathcal{C})}{N})} \tag{7}$$

where

$$E_{\mathrm{r}}(R) = \max_{0 \leq \rho \leq 1} (E_0(\rho) - \rho R) \tag{8}$$

$$E_0(\rho) \triangleq -\log_2 \left\{ \sum_y \left[ \frac{1}{2} p(y|0)^{\frac{1}{1+\rho}} + \frac{1}{2} p(y|1)^{\frac{1}{1+\rho}} \right]^{1+\rho} \right\}. \tag{9}$$

$E_{\mathrm{r}}$ is the random coding error exponent [10], $R \triangleq \frac{K}{N}$ designates the code rate in bits per channel use, and

$$\alpha(\mathcal{C}) \triangleq \max_{1 \leq l \leq N} \frac{A_l}{2^{-N(1-R)} \binom{N}{l}}. \tag{10}$$

In the RHS of (10), $\{A_l\}$ denotes the distance spectrum of the code. Hence, for fully random block codes, $\alpha(\mathcal{C})$ is equal to 1, and the Shulman-Feder bound (SFB) particularizes to the random coding bound [10]. In general, the parameter $\alpha(\mathcal{C})$ in the SFB (7) measures the maximal ratio of the distance spectrum of a code (or ensemble) and the average distance spectrum which corresponds to fully random block codes of the same block length and rate.

The original proof of the SFB is quite involved. In [21], a simpler proof of the SFB is derived, and by doing so, the simplified proof reproduces the SFB as a particular case of the DS2 bound (see Eq. (4)). In light of the significance of the proof concept to the continuation of our paper, we outline this proof briefly.

Since we deal with linear block codes and the communication channel is memoryless, binary-input output-symmetric channel (MBIOS), one can assume without any loss of generality that the all zero codeword $\mathbf{c}_0$ is the transmitted vector. In order to facilitate the expression of the upper bound (6) in terms of distance spectrum of the block code $\mathcal{C}$, we consider here the case where the un-normalized tilting measure $G_N^0(\mathbf{y})$ can be expressed in the following product form:

$$G_N^0(\mathbf{y}) = \prod_{i=1}^{N} g(y_i) \tag{11}$$

where $g$ is an arbitrary non-negative scalar function, and the channel is by assumption MBIOS, so that the transition probability measure is expanded in the product form

$$p_N(\mathbf{y}|\mathbf{c}_{m'}) = \prod_{i=1}^{N} p(y_i|c_{m',i}) \tag{12}$$

where $\mathbf{c}_{m'} = (c_{m',1}, \ldots, c_{m',N})$. Hence, the upper bound on the conditional ML decoding error probability given in (6) can be rewritten as

$$\begin{aligned}
P_{\rm e} &= P_{{\rm e}|0} \\
&\leq \left(\sum_y g(y)\, p(y|0)\right)^{N(1-\rho)} \\
&\quad \cdot \left\{\sum_{l=1}^{N} A_l \left(\sum_y g(y)^{1-\frac{1}{\rho}} p(y|0)\right)^{N-l} \left(\sum_y g(y)^{1-\frac{1}{\rho}} p(y|0)^{1-\lambda} p(y|1)^{\lambda}\right)^{l}\right\}^{\rho} \quad \begin{array}{c}\lambda \geq 0,\\ 0 \leq \rho \leq 1\end{array} \\
&\leq \left(\max_{0<l\leq N} \frac{A_l}{2^{-N(1-R)}\binom{N}{l}}\right)^{\rho} \left(\sum_y g(y)\, p(y|0)\right)^{N(1-\rho)} 2^{-N(1-R)\rho} \\
&\quad \cdot \left\{\sum_y g(y)^{1-\frac{1}{\rho}} p(y|0) + \sum_y g(y)^{1-\frac{1}{\rho}} p(y|0)^{1-\lambda} p(y|1)^{\lambda}\right\}^{N\rho}.
\end{aligned} \tag{13}$$

By setting

$$g(y) = \left[\frac{1}{2} p(y|0)^{\frac{1}{1+\rho}} + \frac{1}{2} p(y|1)^{\frac{1}{1+\rho}}\right]^{\rho} p(y|0)^{-\frac{\rho}{1+\rho}}, \quad \lambda = \frac{1}{1+\rho} \tag{14}$$

and using the symmetry of the channel (where $p(y|0) = p(-y|1)$), the SFB follows readily.

## 2.3 The Tangential-Sphere Bound

The tangential-sphere bound (TSB) on the block error probability of ML decoding was derived by Poltyrev (see [12, 14]), and it applies to the case where the transmission takes place over an AWGN channel, and the modulated signals have constant energy.

Consider a binary linear block code $\mathcal{C}$ with length $N$, and suppose that the modulated codewords are transmitted over a binary-input AWGN channel, using an equi-energy modulation (i.e., all the transmitted signals have the same energy, $E_{\rm s}$, and can be therefore represented as points on an $N$-dimensional sphere whose center is located at the origin). The TSB is based on inequality (1), where the chosen volume $\mathcal{R}$ is an $N$-dimensional circular cone, whose vertex is located at the origin and whose central line passes through the origin and point which corresponds to the transmitted signal (see Fig. 1). This circular cone has a half angle of $\theta$ and a radius $r$, and the optimization is carried over $r$ ($r$ and $\theta$ are related, as shown in Fig. 1). Let us designate the circular cone in Fig. 1 by $C_N(\theta)$.

Let $\mathbf{z} = (z_1, z_2, \ldots, z_N)$ designate an $n$-dimensional noise vector which corresponds to $N$ orthogonal projections of the AWGN. Let $z_1$ be the radial component of $\mathbf{z}$ (see Fig. 1), so the other $N-1$ components of $\mathbf{z}$ are orthogonal to its radial component. Since $\mathbf{z}$ is a Gaussian vector and its components are un-correlated, then the $N$ components of $\mathbf{z}$ are i.i.d., and each component has a zero mean and variance $\sigma^2 = \frac{N_0}{2}$.

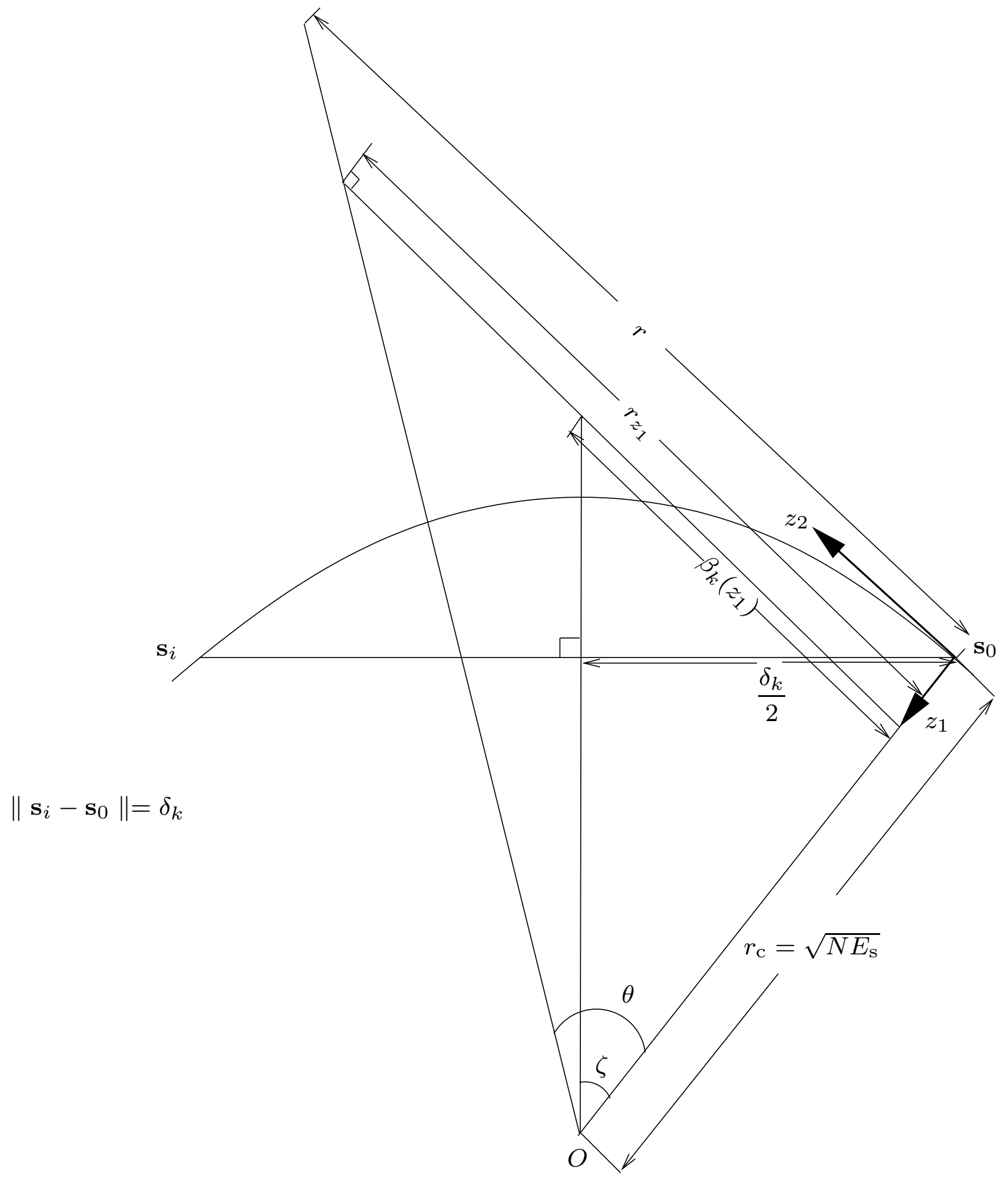

Figure 1: The geometric interpretation of the TSB.

Let $\mathbf{y} = \mathbf{s}_0 + \mathbf{z}$ be the received vector at the output of the binary-input AWGN channel. Given the value of the radial component of the noise vector, $z_1$, the conditional error probability satisfies

$$\text{Prob}\left(E(z_1)|\, z_1\right) \leq \text{Prob}\left(E(z_1),\ \mathbf{y} \in C_N(\theta) \mid z_1\right) + \text{Prob}\left(\mathbf{y} \notin C_N(\theta) \mid z_1\right) \tag{15}$$

and from the union bound

$$\text{Prob}\left(E(z_1),\ \mathbf{y} \in C_N(\theta) \mid z_1\right) \leq \sum_k A_k \text{ Prob}\left(E_k(z_1),\ \mathbf{y} \in C_N(\theta) \mid z_1\right) \tag{16}$$

where $A_k$ designates the number of the constant-energy signals ($\mathbf{s}_i$) in the considered signal set so that their Euclidean distance from the transmitted signal ($\mathbf{s}_0$) is $\delta_k$. We note that for BPSK modulated signals where $\mathbf{s} = (2\mathbf{c} - \underline{1})\sqrt{E_\mathrm{s}}$, the Euclidean distance between the two signals $\mathbf{s}_i$ and $\mathbf{s}_0$ is directly linked to the Hamming weight of the codeword $\mathbf{c}_i$. Let the Hamming distance between the two codewords be $k$, i.e., $w_\mathrm{H}(\mathbf{c}_i) = k$, then the Euclidean distance between the two BPSK modulated signals is equal to $\delta_k = 2\sqrt{kE_\mathrm{s}}$. In the latter case, $A_k$ is the number of codewords of the code $\mathcal{C}$ with Hamming weight $k$ (i.e., $\{A_k\}$ is the distance spectrum of the linear code $\mathcal{C}$).

The combination of Eqs. (15) and (16) gives

$$\begin{aligned} \text{Prob}\left(E(z_1)|\ z_1\right) \leq & \sum_k \Big\{A_k \ \text{Prob}\left(E_k(z_1),\ \mathbf{y} \in C_N(\theta) \mid z_1\right)\Big\} \\ & + \text{Prob}\left(\mathbf{y} \notin C_N(\theta) \mid z_1\right). \end{aligned} \tag{17}$$

From Fig. 1, we obtain the following equalities:

$$\begin{cases} r_{z_1} &= \left(1 - \frac{z_1}{\sqrt{NE_\mathrm{s}}}\right) r \\ \beta_k(z_1) &= \frac{r_{z_1}}{\sqrt{1 - \frac{\delta_k^2}{4NE_\mathrm{s}}}} \frac{\delta_k}{2r}. \end{cases} \tag{18}$$

and let

$$\alpha_k \triangleq r\sqrt{1 - \frac{\delta_k^2}{4NE_\mathrm{s}}}.$$

The TSB on the block error probability under ML decoding depends on the distance spectrum $\{A_l\}_{l=0}^N$ of the block code $\mathcal{C}$, and by invoking the smoothing theorem w.r.t. $z_1$, the bound finally admits the following form:

$$P_\mathrm{e} \leq \int_{-\infty}^{\infty} \frac{dz_1}{\sqrt{2\pi}\sigma} e^{-\frac{z_1^2}{2\sigma^2}} \left\{ \begin{array}{l} \displaystyle\sum_{k: \frac{\delta_k}{2} \leq \alpha_k} \left\{ A_k \int_{\beta_k(z_1)}^{r_{z_1}} \frac{1}{\sqrt{2\pi}\sigma} e^{-\frac{z_2^2}{2\sigma^2}} \ \bar{\gamma}\left(\frac{N-2}{2}, \frac{r_{z_1}^2 - z_2^2}{2\sigma^2}\right) dz_2 \right\} \\ \qquad + 1 - \bar{\gamma}\left(\frac{N-1}{2}, \frac{r_{z_1}^2}{2\sigma^2}\right) \end{array} \right\} \tag{19}$$

where

$$\bar{\gamma}(a, x) \triangleq \frac{1}{\Gamma(a)} \int_0^x t^{a-1} e^{-t} dt, \quad a, x > 0 \tag{20}$$

denotes the normalized incomplete Gamma function. In order to get the tightest bound within the form (19), one needs to optimize the radius $r$. This optimization is carried by setting to zero the derivative of the RHS of (19) w.r.t. $r$, which yields the following optimization equations ([14]):

$$\begin{cases} \displaystyle\sum_{k: \frac{\delta_k}{2} < \alpha_k} A_k \int_0^{\theta_k} \sin^{N-3}\phi \ d\phi = \frac{\sqrt{\pi}\ \Gamma(\frac{N-2}{2})}{\Gamma(\frac{N-1}{2})} \\ \theta_k = \cos^{-1}\left( \frac{\delta_k}{2r} \frac{1}{\sqrt{1 - \frac{\delta_k^2}{4NE_\mathrm{s}}}} \right). \end{cases} \tag{21}$$

In [15, Appendix B], Sason and Shamai prove the existence and uniqueness of the solution to the optimization equation (21), and introduce a simple algorithm for solving this equation numerically. In order to derive an upper bound on the bit error probability, let $A_{w,k}$ designate the corresponding coefficient in the IOWEF which is number of codewords which are encoded by information bits whose number of ones is equal to $w$ (where $0 \leq w \leq NR$) and whose Hamming weights (after encoding) are equal to $k$, and define

$$A'_k \triangleq \sum_{w=1}^{NR} \left(\frac{w}{NR}\right) A_{w,k}, \quad k = 0, \ldots, N. \tag{22}$$

In [15, Appendix C], Sason and Shamai derive an upper bound on the bit error probability by replacing the distance spectrum $\{A_k\}$ in (19) and (21) with the sequence $\{A'_k\}$, and show some properties on the resulting bound on the bit error probability.

# 3 Improved Upper Bounds

## 3.1 Upper Bound on the Block Error Probability

It is well known that at rates below the channel capacity, the block error probability of the ensemble of fully random block codes vanishes exponentially with the block length. In the random coding setting, the TSB [14] fails to reproduce the random coding exponent, while the SFB [20] particularizes to the 1965 Gallager bound for random codes, and hence, the SFB reproduces the random coding exponent. The SFB is therefore advantageous over the TSB in the random coding setting when we let the block length be sufficiently large. Equations (7) and (10) imply that for specific linear codes (or ensembles), the tightness of the SFB depends on the maximal ratio between the distance spectrum of the code (or the average distance spectrum of the ensemble) and the average distance spectrum of fully random block codes of the same length and rate which has a binomial distribution.

In order to tighten the SFB bound for linear block codes, Miller and Burshtein [13] suggested to partition the original linear code $\mathcal{C}$ into two subcodes, namely $\mathcal{C}'$ and $\mathcal{C}''$; the subcode $\mathcal{C}'$ contains the all-zero codeword and all the codewords with Hamming weights of $l \in \mathcal{U} \subseteq \{1, 2, ..., N\}$, while $\mathcal{C}''$ contains the other codewords which have Hamming weights of $l \in \mathcal{U}^{\mathrm{c}} = \{1, 2, ..., N\} \setminus \mathcal{U}$ and the all-zero codeword. From the symmetry of the channel, the union bound provides the following upper bound on the ML decoding error probability:

$$P_{\mathrm{e}} = P_{\mathrm{e}|0} \leq P_{\mathrm{e}|0}(\mathcal{C}') + P_{\mathrm{e}|0}(\mathcal{C}'') \tag{23}$$

where $P_{\mathrm{e}|0}(\mathcal{C}')$ and $P_{\mathrm{e}|0}(\mathcal{C}'')$ designate the conditional ML decoding error probabilities of $\mathcal{C}'$ and $\mathcal{C}''$, respectively, given that the all zero codeword is transmitted. We note that although the code $\mathcal{C}$ is linear, its two subcodes $\mathcal{C}'$ and $\mathcal{C}''$ are in general *non-linear*. One can rely on different upper bounds on the conditional error probabilities $P_{\mathrm{e}|0}(\mathcal{C}')$ and $P_{\mathrm{e}|0}(\mathcal{C}'')$, i.e., we may bound $P_{\mathrm{e}|0}(\mathcal{C}')$ by the SFB, and rely on an alternative approach to obtain an upper bound on $P_{\mathrm{e}|0}(\mathcal{C}'')$. For example, if we consider the binary-input AWGN channel, then the TSB (or even union bounds) may be used in order to obtain an upper bound on the conditional error probability $P_{\mathrm{e}|0}(\mathcal{C}'')$ which corresponds to the subcode $\mathcal{C}''$. In order to obtain the tightest bound in this approach, one should look for an optimal partitioning of the original code $\mathcal{C}$ into two sub-codes, based on the distance spectrum of $\mathcal{C}$. The solution of the problem is quite tedious, because in general, if the subset $\mathcal{U}$ can be an arbitrary subset of the set of integers $\{1, \ldots, N\}$, then one has to compare $\sum_{i=0}^{N} \binom{N}{i} = 2^N$ different possibilities for $\mathcal{U}$. However, we may use practical optimization schemes to obtain good results which may improve the tightness of both the SFB and TSB.

An easy way to make an efficient partitioning of a linear block code $\mathcal{C}$ is to compare its distance spectrum (or the average distance spectrum for an ensemble of linear codes) with the average distance spectrum of the ensemble of fully random block codes of the same rate and block length. Let us designate the average distance spectrum of the ensemble of fully random block codes of length $N$ and rate $R$ by

$$B_l \triangleq 2^{-N(1-R)} \binom{N}{l} \qquad l = 0, 1, \ldots, N. \tag{24}$$

Then, it is suggested to partition $\mathcal{C}$ in a way so that all the codewords with Hamming weight $l$ for which $\frac{A_l}{B_l}$ is greater than some threshold (which should be larger than 1 but close to it) are associated with $\mathcal{C}''$, and the other codewords are associated with $\mathcal{C}'$. The following algorithm is suggested for the calculation of the upper bound on the block error probability under ML decoding:

**Algorithm 1**

**1.** Set

$$\mathcal{U} = \Phi, \quad \mathcal{U}^{\text{c}} = \{1, 2, ...N\}, \quad l = 1$$

where $\Phi$ designates an empty set, and set the initial value of the upper bound to be 1.

**2.** Compute the ratio $\frac{A_l}{B_l}$ where $\{A_l\}$ is the distance spectrum of the binary linear block code (or the average distance of an ensemble of such codes), and $\{B_l\}$ is the binomial distribution introduced in (24).

**3.** If this ratio is smaller than some threshold (where the value of the threshold is typically set to be slightly larger than 1), then the element $l$ is added to the set $\mathcal{U}$, i.e.,

$$\mathcal{U} := \mathcal{U} + \{l\}, \quad \mathcal{U}^{\text{c}} := \mathcal{U}^{\text{c}} \setminus \{l\}.$$

**4.** Update correspondingly the upper bound in the RHS of (23) (we will derive later the appropriate upper bounds on $P_{\text{e}|0}(\mathcal{C}')$ and $P_{\text{e}|0}(\mathcal{C}'')$.

**5.** Set the bound to be the minimum between the RHS from Step 4 and its previous value.

**6.** Set $l = l + 1$ and go to Step 2.

**7.** The algorithm terminates when $l$ gets the value $N$ (i.e., the block length of the code) or actually, the maximal value of $l$ for which $A_l$ does not vanish.[1]

[1]The number of steps can be reduced by factor of 2 for binary linear codes which contain the all-ones codeword (hence maintain the property $A_l = A_{N-l}$). For such codes, the update equation in Step 3 becomes: $\mathcal{U} := \mathcal{U} + \{l, N-l\}, \quad \mathcal{U}^{\text{c}} := \mathcal{U}^{\text{c}} - \{l, N-l\}$ and the algorithm terminates when $l$ gets the value $\lceil \frac{N}{2} \rceil$.

Fig. 2(a) shows a plot of the ratio $\frac{A_l}{B_l}$ as a function of $\delta \triangleq \frac{l}{N}$ for an ensemble of uniformly interleaved turbo-random codes. The calculation of the average distance spectrum of these ensemble relies on the results of Soljanin and Urbanke in [22]. From the discussion above, it is clear that the combination of the SFB with another upper bound has the potential to tighten the overall upper bound on the ML decoding probability; this improvement is expected to be especially pronounced for ensembles whose average distance spectrum resembles the binomial distribution of fully random block codes over a relatively large range of Hamming weights, but deviates significantly from the binomial distribution for relatively low and large Hamming weights (e.g., ensembles of uniformly interleaved turbo codes possess this property, as indicated in [15, Section 4]). This bounding technique was successfully applied by Miller and Burshtein [13] and also by Sason and Urbanke [18] to ensembles of regular LDPC codes where the SFB was combined with union bounds. If the range of Hamming weights where the average distance spectrum of an ensemble resembles the binomial distribution is relatively large, then according to the above algorithm, one would expect that $\mathcal{C}'$ typically contains a very large fraction of the overall number of the codewords of a code from this ensemble. Hence, in order to obtain an upper bound on $P_{\mathrm{e}|0}(\mathcal{C}'')$, where $\mathcal{C}''$ is expected to contain a rather small fraction of the codewords in $\mathcal{C}$, we may use a simple bound such as the union bound while expecting not to pay a significant penalty in the tightness of the overall bound on the decoding error probability ($P_{\mathrm{e}}$).

The following bound introduced in Theorem 1 is derived as a particular case of the DS2 bound [8]. The beginning of its derivation is similar to the steps in [21, Section 4A], but we later deviate from the analysis there in order to modify the SFB. We finally obtain a tighter version of this bound.

**Theorem 1 (Modified Shulman and Feder Bound).** Let $\mathcal{C}$ be a binary linear block code of length $N$ and rate $R$, and let $\{A_l\}$ designate its distance spectrum. Let this code be partitioned into two subcodes, $\mathcal{C}'$ and $\mathcal{C}''$, where $\mathcal{C}'$ contains the all-zero codeword and all the other codewords of $\mathcal{C}$ whose Hamming weights are in an arbitrary set $\mathcal{U} \subseteq \{1, 2, , \ldots, N\}$; the second subcode $\mathcal{C}''$ contains the all-zero codeword and the other codewords of $\mathcal{C}$ which are not included in $\mathcal{C}'$. Assume that the communication takes place over a memoryless binary-input output-symmetric (MBIOS) channel with transition probability measure $p(y|x),\ x \in \{0,1\}$. Then, the block error probability of $\mathcal{C}$ under ML decoding is upper bounded by

$$P_{\mathrm{e}} \leq P_{\mathrm{e}|0}(\mathcal{C}') + P_{\mathrm{e}|0}(\mathcal{C}'')$$

where

$$P_{\mathrm{e}|0}(\mathcal{C}') \leq \mathrm{SFB}(\rho) \cdot \left[ \sum_{l \in \mathcal{U}} \binom{N}{l} \left( \frac{A(\rho)}{A(\rho)+B(\rho)} \right)^{l} \left( \frac{B(\rho)}{A(\rho)+B(\rho)} \right)^{N-l} \right]^{\rho}, \quad 0 \leq \rho \leq 1 \quad (25)$$

$$A(\rho) \triangleq \sum_{y} \left\{ [p(y|0)p(y|1)]^{\frac{1}{1+\rho}} \left[ \frac{1}{2} p(y|0)^{\frac{1}{1+\rho}} + \frac{1}{2} p(y|1)^{\frac{1}{1+\rho}} \right]^{\rho-1} \right\} \quad (26)$$

$$B(\rho) \triangleq \sum_{y} \left\{ p(y|0)^{\frac{2}{1+\rho}} \left[ \frac{1}{2} p(y|0)^{\frac{1}{1+\rho}} + \frac{1}{2} p(y|1)^{\frac{1}{1+\rho}} \right]^{\rho-1} \right\}. \quad (27)$$

The multiplicative term, $\mathrm{SFB}(\rho)$, in the RHS of (25) designates the conditional Shulman-Feder upper bound of the subcode $\mathcal{C}'$ given the transmission of the all-zero codeword, i.e.,

$$\mathrm{SFB}(\rho) = 2^{-N\left(E_0(\rho)-\rho(R+\frac{\log(\alpha(\mathcal{C}'))}{N})\right)}, \quad 0 \le \rho \le 1 \tag{28}$$

and $E_0$ is introduced in (9). An upper bound on the conditional block error probability for the subcode $\mathcal{C}''$, $P_{\mathrm{e}|0}(\mathcal{C}'')$, can be either a standard union bound or any other bound.

*Proof.* Since the block code $\mathcal{C}$ is linear and the channel is MBIOS, the conditional block error probability of $\mathcal{C}$ is independent of the transmitted codeword. Hence, the union bound gives the following upper bound on the block error probability: $P_{\mathrm{e}} \le P_{\mathrm{e}|0}(\mathcal{C}') + P_{\mathrm{e}|0}(\mathcal{C}'')$.

In order to prove the theorem, we derive an upper bound on $P_{\mathrm{e}|0}(\mathcal{C}')$. Let $\{A_l(\mathcal{C}')\}$ denote the weight spectrum of the subcode $\mathcal{C}'$, and let $G_N(\mathbf{y})$ be an arbitrary non-negative function of the received vector $\mathbf{y} = (y_1, y_2, \ldots, y_N)$ where this function is assumed to be expressible in the product form (11). Then, we get from (6) and (11) the following upper bound on the conditional ML decoding error probability of the subcode $\mathcal{C}'$:

$$\begin{aligned}
P_{\mathrm{e}|0}(\mathcal{C}') &\le \left(\sum_y g(y)\, p(y|0)\right)^{N(1-\rho)} \\
&\quad \cdot \left\{\sum_l A_l(\mathcal{C}') \left(\sum_y g(y)^{1-\frac{1}{\rho}} p(y|0)\right)^{N-l} \left(\sum_y g(y)^{1-\frac{1}{\rho}} p(y|0)^{1-\lambda} p(y|1)^{\lambda}\right)^{l}\right\}^{\rho} \quad \lambda \ge 0,\ 0 \le \rho \le 1 \\
&= \left(\sum_y g(y)\, p(y|0)\right)^{N(1-\rho)} 2^{-N(1-R)\rho} \\
&\quad \cdot \left\{\sum_{l \in \mathcal{U}} \left(\frac{A_l}{2^{-N(1-R)}\binom{N}{l}}\right) \binom{N}{l} \left(\sum_y g(y)^{1-\frac{1}{\rho}} p(y|0)\right)^{N-l} \right. \\
&\qquad \left. \left(\sum_y g(y)^{1-\frac{1}{\rho}} p(y|0)^{1-\lambda} p(y|1)^{\lambda}\right)^{l}\right\}^{\rho} \\
&\le \left(\max_{l \in \mathcal{U}} \frac{A_l}{2^{-N(1-R)}\binom{N}{l}}\right)^{\rho} \left(\sum_y g(y)\, p(y|0)\right)^{N(1-\rho)} 2^{-N(1-R)\rho} \\
&\quad \cdot \left\{\sum_{l \in \mathcal{U}} \binom{N}{l} \left(\sum_y g(y)^{1-\frac{1}{\rho}} p(y|0)\right)^{N-l} \left(\sum_y g(y)^{1-\frac{1}{\rho}} p(y|0)^{1-\lambda} p(y|1)^{\lambda}\right)^{l}\right\}^{\rho}.
\end{aligned} \tag{29}$$

The transition in the first equality above follows since $A_l(\mathcal{C}') \equiv 0$ for $l \notin \mathcal{U}$, and $A_l(\mathcal{C}')$ coincide with the distance spectrum of the code $\mathcal{C}$ for all $l \in \mathcal{U}$. Note that (29) is a tighter version of the bound in [21, Eq. (32)]. The difference between the modified and the original bounds is that in the former, we only sum over the indices $l \in \mathcal{U}$ while in the latter, we sum over the whole set of indices, i.e., $l \in \{1, 2, \ldots, N\}$. By setting the tilting measure in (14), the symmetry of the MBIOS channel gives the equality

$$\sum_y g(y) p(y|0) = \sum_y \left[\frac{1}{2} p(y|0)^{\frac{1}{1+\rho}} + \frac{1}{2} p(y|1)^{\frac{1}{1+\rho}}\right]^{\rho+1} \tag{30}$$

and from (26) and (27)

$$\begin{aligned}
&\sum_y p(y|0)^{1-\lambda} p(y|1)^{\lambda} g(y)^{1-\frac{1}{\rho}} \\
&= \sum_y \left\{ [p(y|0)p(y|1)]^{\frac{1}{1+\rho}} \left[ \frac{1}{2} p(y|0)^{\frac{1}{1+\rho}} + \frac{1}{2} p(y|1)^{\frac{1}{1+\rho}} \right]^{\rho-1} \right\} \\
&= A(\rho) \qquad (31)
\end{aligned}$$

$$\begin{aligned}
&\sum_y p(y|0) g(y)^{1-\frac{1}{\rho}} \\
&= \sum_y \left\{ p(y|0)^{\frac{2}{1+\rho}} \left[ \frac{1}{2} p(y|0)^{\frac{1}{1+\rho}} + \frac{1}{2} p(y|1)^{\frac{1}{1+\rho}} \right]^{\rho-1} \right\} \\
&= B(\rho) \qquad (32)
\end{aligned}$$

where the RHS of (31) and (32) are obtained by setting $\lambda = \frac{1}{1+\rho}$. Finally, based on (14) and the symmetry of the channel, one can verify that

$$\sum_y g(y) p(y|0) = \frac{A(\rho) + B(\rho)}{2}. \qquad (33)$$

Substituting (30)–(33) into (29) gives the following conditional upper bound on the ML decoding error probability of the subcode $\mathcal{C}'$:

$$P_{\mathrm{e}|0}(\mathcal{C}') \leq \alpha(\mathcal{C}')^{\rho} \left( \frac{A(\rho) + B(\rho)}{2} \right)^{N(1-\rho)} 2^{-N(1-R)\rho} \cdot \left( \sum_{l \in \mathcal{U}} \binom{N}{l} A^l(\rho) B^{N-l}(\rho) \right)^{\rho} \qquad (34)$$

where we use the notation

$$\alpha(\mathcal{C}') \triangleq \max_{l \in \mathcal{U}} \frac{A_l}{2^{-N(1-R)} \binom{N}{l}}.$$

The latter parameter measures by how much the (expected) number of codewords in the subcode $\mathcal{C}'$ deviates from the binomial distribution which characterizes the average distance spectrum of the ensemble of fully random block codes of length $N$ and rate $R$. By straightforward algebra, we obtain that

$$\begin{aligned}
P_{\mathrm{e}|0}(\mathcal{C}') &\leq \alpha(\mathcal{C}')^{\rho} \left( \frac{A(\rho) + B(\rho)}{2} \right)^{N} 2^{-N(1-R)\rho} \left( \frac{1}{2} \right)^{-N\rho} \\
&\quad \cdot \left[ \sum_{l \in \mathcal{U}} \binom{N}{l} \left( \frac{A(\rho)}{A(\rho) + B(\rho)} \right)^{l} \left( \frac{B(\rho)}{A(\rho) + B(\rho)} \right)^{N-l} \right]^{\rho} \\
&= \alpha(\mathcal{C}')^{\rho} \left( \frac{A(\rho) + B(\rho)}{2} \right)^{N} 2^{NR\rho} \left[ \sum_{l \in \mathcal{U}} \binom{N}{l} \left( \frac{A(\rho)}{A(\rho) + B(\rho)} \right)^{l} \left( \frac{B(\rho)}{A(\rho) + B(\rho)} \right)^{N-l} \right]^{\rho} \\
&= \mathrm{SFB}(\rho) \cdot \left[ \sum_{l \in \mathcal{U}} \binom{N}{l} \left( \frac{A(\rho)}{A(\rho) + B(\rho)} \right)^{l} \left( \frac{B(\rho)}{A(\rho) + B(\rho)} \right)^{N-l} \right]^{\rho}, \quad 0 \leq \rho \leq 1. \qquad (35)
\end{aligned}$$

The second equality follows from (28) and (9), and since

$$\begin{aligned} E_0(\rho) &\triangleq -\log_2 \left\{ \sum_y \left[ \frac{1}{2} p(y|0)^{\frac{1}{1+\rho}} + \frac{1}{2} p(y|1)^{\frac{1}{1+\rho}} \right]^{1+\rho} \right\} \\ &= -\log_2 \left( \frac{A(\rho)+B(\rho)}{2} \right). \end{aligned} \quad (36)$$

This concludes the proof of the theorem. □

*Discussion:* The improvement of the bound introduced in Theorem 1 over the standard combination of the SFB and the union bound [13, 18] stems from the introduction of the factor which multiplies $\text{SFB}(\rho)$ in the RHS of (25); this multiplicative term cannot exceed 1 since

$$\begin{aligned} &\sum_{l \in \mathcal{U}} \binom{N}{l} \left( \frac{A(\rho)}{A(\rho)+B(\rho)} \right)^l \left( \frac{B(\rho)}{A(\rho)+B(\rho)} \right)^{N-l} \\ &\leq \sum_{l=0}^{N} \binom{N}{l} \left( \frac{A(\rho)}{A(\rho)+B(\rho)} \right)^l \left( \frac{B(\rho)}{A(\rho)+B(\rho)} \right)^{N-l} = 1. \end{aligned}$$

This multiplicative factor which appears in the new bound is useful for finite-length codes with small to moderate block lengths. The upper bound (25) on $P_{\mathrm{e}|0}(\mathcal{C}')$ is clearly at least as tight as the corresponding conditional SFB. We refer to the upper bound (25) as the modified SFB (MSFB). The conditional block error probability of the subcode $\mathcal{C}''$, given that the all-zero codeword is transmitted, can be bounded by a union bound or any improved upper bound conditioned on the transmission of the all-zero codeword (note that the subcode $\mathcal{C}''$ is in general a non-linear code). In general, one is looking for an appropriate balance between the two upper bounds on $P_{\mathrm{e}|0}^{(1)}$ and $P_{\mathrm{e}|0}^{(2)}$ (see Algorithm 1). The improvement that is achieved by using the MSFB instead of the corresponding SFB is exemplified in Section 5 for ensembles of uniformly interleaved turbo-Hamming codes.

## 3.2 Upper Bounds on Bit Error Probability

Let $\mathcal{C}$ be a binary linear block code whose transmission takes place over an arbitrary MBIOS channel, and let $P_{\mathrm{b}}$ designate the bit error probability of $\mathcal{C}$ under ML decoding. In [16, Appendix A], Sason and Shamai derived an upper bound on the bit error probability of systematic, binary linear block codes which are transmitted over fully interleaved fading channels with perfect channel state information at the receiver. Here we generalize the result of [16] for arbitrary MBIOS channels. In order to derive the desired upper bound we use the following lemma due to Divsalar [6], and provide a simplified proof to this lemma:

**Lemma 1.** [6, Section III.C] Let $\mathcal{C}$ be a binary block code of dimension $K$ whose transmission takes place over an MBIOS channel. Let $\mathcal{C}(w)$ designate a sub-code of $\mathcal{C}$ which includes the all-zero codeword and all the codewords of $\mathcal{C}$ which are encoded by *information bits* whose Hamming weight is $w$. Then the conditional bit error probability of $\mathcal{C}$ under ML decoding, given that the all-zero codeword is transmitted, is upper bounded by

$$P_{\mathrm{b}|0} \leq \sum_{\mathbf{y}} p_N(\mathbf{y}|\mathbf{0})^{1-\lambda\rho} \left\{ \sum_{w=1}^{K} \left(\frac{w}{K}\right) \sum_{\substack{\mathbf{c} \in \mathcal{C}(w) \\ \mathbf{c} \neq \mathbf{0}}} p_N(\mathbf{y}|\mathbf{c})^{\lambda} \right\}^{\rho}, \qquad \lambda > 0, \;\; 0 \leq \rho \leq 1. \tag{37}$$

We introduce here a somewhat simpler proof than in [6].

*Proof.* The conditional bit error probability under ML decoding admits the form

$$P_{\mathrm{b}|0} = \sum_{\mathbf{y}} \left(\frac{w_0(\mathbf{y})}{K}\right) p_N(\mathbf{y}|\mathbf{0}) \tag{38}$$

where $w_0(\mathbf{y}) \in \{0, 1, ..., K\}$ designates the weight of the information bits in the decoded codeword, given the all-zero codeword is transmitted and the received vector is $\mathbf{y}$. In particular, if the received vector $\mathbf{y}$ is included in the decision region of the all-zero codeword, then $w_0(\mathbf{y}) = 0$. The following inequalities hold:

$$\begin{aligned}
\frac{w_0(\mathbf{y})}{K} &\leq \left(\frac{w_0(\mathbf{y})}{K}\right)^{\rho}, \qquad 0 \leq \rho \leq 1 \\
&\stackrel{\text{(a)}}{\leq} \left\{ \left(\frac{w_0(\mathbf{y})}{K}\right) \sum_{\substack{\mathbf{c} \in \mathcal{C}(w_0(\mathbf{y})) \\ \mathbf{c} \neq \mathbf{0}}} \left[\frac{p_N(\mathbf{y}|\mathbf{c})}{p_N(\mathbf{y}|\mathbf{0})}\right]^{\lambda} \right\}^{\rho} \qquad \lambda \geq 0 \\
&\leq \left\{ \sum_{w=1}^{K} \left(\frac{w}{K}\right) \sum_{\substack{\mathbf{c} \in \mathcal{C}(w) \\ \mathbf{c} \neq \mathbf{0}}} \left[\frac{p_N(\mathbf{y}|\mathbf{c})}{p_N(\mathbf{y}|\mathbf{0})}\right]^{\lambda} \right\}^{\rho}.
\end{aligned} \tag{39}$$

Inequality (a) holds since the received vector $\mathbf{y}$ falls in the decision region of a codeword $\tilde{\mathbf{c}}$ which is encoded by information bits of total Hamming weight $w_0(\mathbf{y})$; hence, the quotient $\frac{p_N(\mathbf{y}|\tilde{\mathbf{c}})}{p_N(\mathbf{y}|\mathbf{0})}$ is larger than 1 while the other terms in the sum are simply non-negative. The third inequality holds because of adding non-negative terms to the sum. The lemma follows by substituting (39) into the RHS of (38). □

**Theorem 2. (The SFB Version on the BER)** Let $\mathcal{C}$ be a binary linear block code of length $N$ and dimension $K$, and assume that the transmission of the code takes place over an MBIOS channel. Let $A_{w,l}$ designate the number of codewords in $\mathcal{C}$ which are encoded by information bits whose Hamming weight is $w$ and their Hamming weight after encoding is $l$. Then, the bit error probability of $\mathcal{C}$ under ML decoding is upper bounded by

$$P_{\mathrm{b}} \leq 2^{-NE_{\mathrm{r}}(R + \frac{\log \alpha_{\mathrm{b}}(\mathcal{C})}{N})} \tag{40}$$

where $R = \frac{K}{N}$ is the code rate of $\mathcal{C}$, and

$$\alpha_{\mathrm{b}}(\mathcal{C}) \triangleq \max_{0<l\leq N} \frac{A'_l}{2^{-N(1-R)} \binom{N}{l}}, \qquad A'_l \triangleq \sum_{w=1}^{K} \left(\frac{w}{K}\right) A_{w,l}.$$

*Proof.* Due to the linearity of the code $\mathcal{C}$ and the symmetry of the channel, the conditional bit error probability of the code is independent on the transmitted codeword; hence, without any loss of generality, it is assumed that the all-zero codeword is transmitted. From (37), the following upper bound on the bit error probability of $\mathcal{C}$ follows:

$$\begin{aligned} P_{\mathrm{b}} = P_{\mathrm{b}|0} &\leq \sum_{\mathbf{y}} p_N(\mathbf{y}|\mathbf{0})^{1-\lambda\rho} \left\{ \sum_{w=1}^{K} \left(\frac{w}{K}\right) \sum_{\substack{\mathbf{c} \in \mathcal{C}(w) \\ \mathbf{c} \neq \mathbf{0}}} p_N(\mathbf{y}|\mathbf{c})^{\lambda} \right\}^{\rho}, \qquad \lambda > 0, \ \ 0 \leq \rho \leq 1 \\ &= \sum_{\mathbf{y}} \psi_N^0(\mathbf{y}) \left\{ \psi_N^0(\mathbf{y})^{-\frac{1}{\rho}} p_N(\mathbf{y}|\mathbf{0})^{\frac{1}{\rho}} \sum_{w=1}^{K} \left(\frac{w}{K}\right) \sum_{\substack{\mathbf{c} \in \mathcal{C}(w) \\ \mathbf{c} \neq \mathbf{0}}} \left[\frac{p_N(\mathbf{y}|\mathbf{c})}{p_N(\mathbf{y}|\mathbf{0})}\right]^{\lambda} \right\}^{\rho} \end{aligned} \tag{41}$$

where $\psi_N^0$ is an arbitrary probability tilting measure. By invoking Jensen inequality in the RHS of (41) and replacing $\psi_N^0(\mathbf{y})$ with the un-normalized tilting measure $G_N^0(\mathbf{y})$ which appears in the RHS of (5), the upper bound in (41) transforms to

$$\begin{aligned} P_{\mathrm{b}|0} &\leq \left( \sum_{\mathbf{y}} G_N^0(\mathbf{y})\, p_N(\mathbf{y}|\mathbf{0}) \right)^{1-\rho} \\ &\quad \cdot \left\{ \sum_{w=1}^{k} \left(\frac{w}{k}\right) \sum_{\substack{\mathbf{c} \in \mathcal{C}(w) \\ \mathbf{c} \neq \mathbf{0}}} \sum_{\mathbf{y}} p_N(\mathbf{y}|\mathbf{0}) G_N^0(\mathbf{y})^{1-\frac{1}{\rho}} \left[\frac{p_N(\mathbf{y}|\mathbf{c})}{p_N(\mathbf{y}|\mathbf{0})}\right]^{\lambda} \right\}^{\rho}, \ 0 \leq \rho \leq 1, \ \lambda > 0. \end{aligned} \tag{42}$$

We consider an un-normalized tilting measure $G_N^0(\mathbf{y})$ which is expressible in the product form (11). Since the communication channel is MBIOS and $\mathcal{C}$ is a binary linear block code, one obtains the following upper bound on the bit error probability:

$$\begin{aligned} P_{\mathrm{b}|0} &\leq \left( \sum_{y} g(y)\, p(y|0) \right)^{N(1-\rho)} \qquad 0 \leq \rho \leq 1, \ \lambda > 0 \\ &\quad \cdot \left\{ \sum_{w=1}^{K} \left(\frac{w}{K}\right) \sum_{l=0}^{N} A_{w,l} \left( \sum_{y} p(y|0) g(y)^{1-\frac{1}{\rho}} \right)^{N-l} \left( \sum_{y} p(y|1)^{\lambda} p(y|0)^{1-\lambda} g(y)^{1-\frac{1}{\rho}} \right)^{l} \right\}^{\rho} \\ &= \left( \sum_{y} g(y)\, p(y|0) \right)^{N(1-\rho)} \\ &\quad \cdot \left\{ \sum_{l=0}^{N} A'_l \left( \sum_{y} p(y|0) g(y)^{1-\frac{1}{\rho}} \right)^{N-l} \left( \sum_{y} p(y|1)^{\lambda} p(y|0)^{1-\lambda} g(y)^{1-\frac{1}{\rho}} \right)^{l} \right\}^{\rho} \\ &\leq \left( \sum_{y} g(y)\, p(y|0) \right)^{N(1-\rho)} \left( \max_{0 \leq l \leq N} \frac{A'_l}{2^{-N(1-R)} \binom{n}{l}} \right)^{\rho} \cdot 2^{-N(1-R)\rho} \\ &\quad \cdot \left( \sum_{y} p(y|1)^{\lambda} p(y|0)^{1-\lambda} g(y)^{1-\frac{1}{\rho}} + \sum_{y} p(y|1)^{\lambda} p(y|0)^{1-\lambda} g(y)^{1-\frac{1}{\rho}} \right)^{N\rho} \end{aligned} \tag{43}$$

By setting $g(y)$ as in (14), we obtain an upper bound which is the same as the original SFB, except that the distance spectrum $\{A_l\}$ is replaced by $\{A_l{}'\}$. This provides the bound introduced in (40), and concludes the proof of the theorem. □

Similarly to the derivation of the combined upper bound on the block error probability in Theorem 1, we suggest to partition the code into two subcodes in order to get improved upper bounds on the bit error probability; however, since we consider the bit error probability instead of block error probability, the threshold in Algorithm 1 is typically modified to a value which is slightly above $\frac{1}{2}$ (instead of 1). Since the code is linear and the channel is MBIOS, the conditional decoding error probability is independent of the transmitted codeword (so, we assume again that the all-zero codeword is transmitted). By the union bound

$$P_{\mathrm{b}} = P_{\mathrm{b}|0} \leq P_{\mathrm{b}|0}(\mathcal{C}') + P_{\mathrm{b}|0}(\mathcal{C}'') \tag{44}$$

where $P_{\mathrm{b}|0}(\mathcal{C}')$ and $P_{\mathrm{b}|0}(\mathcal{C}'')$ denote the conditional ML decoding bit error probabilities of two disjoint subcodes $\mathcal{C}'$ and $\mathcal{C}''$ which partition the block code $\mathcal{C}$ (except that these two subcodes have the all-zero vector in common), given that the all-zero codeword is transmitted. The construction of the subcodes $\mathcal{C}'$ and $\mathcal{C}''$ is characterized later.

*Upper bound on* $P_{\mathrm{b}|0}(\mathcal{C}')$: Let $A_{w,l}$ designate the number of codewords of Hamming weight $l$ which are encoded by a sequence of information bits of Hamming weight $w$. Similarly to the discussion on the block error probability, we use the bit-error version of the SFB (see Eq. (40)) as an upper bound on $P_{\mathrm{b}|0}(\mathcal{C}')$. From Theorem 2, it follows that the conditional bit error probability of the subcode $\mathcal{C}'$, given that the all-zero codeword is transmitted is upper bounded by

$$P_{\mathrm{b}|0}(\mathcal{C}') \leq 2^{-NE_{\mathrm{r}}\left(R+\frac{\log \alpha_{\mathrm{b}}(\mathcal{C}')}{N}\right)} \tag{45}$$

where

$$\alpha_{\mathrm{b}}(\mathcal{C}') \triangleq \max_{l\in\mathcal{U}} \frac{A_l'(\mathcal{C}')}{B_l} \quad , A_l{}'(\mathcal{C}') \triangleq \begin{cases} \sum_{w=1}^{NR} \left(\frac{w}{NR}\right) A_{w,l} & \text{if } l \in \mathcal{U} \\ 0 & \text{otherwise} \end{cases} \tag{46}$$

and the set $\mathcal{U}$ in (46) stands for an arbitrary subset of $\{1, \ldots, N\}$.

*Upper bound on* $P_{\mathrm{b}|0}(\mathcal{C}'')$: We may bound the conditional bit error probability of the subcode $\mathcal{C}''$, $P_{\mathrm{b}|0}(\mathcal{C}'')$, by an improved upper bound. For the binary-input AWGN, the modified version of the TSB, as shown in [15] is an appropriate bound. This bound is the same as the original TSB in (19), except that the distance spectrum $\{A_l\}$ is replaced by $\{A_l{}'(\mathcal{C}'')\}$ where

$$A_l{}'(\mathcal{C}'') \triangleq \begin{cases} \displaystyle\sum_{w=1}^{NR} \left(\frac{w}{NR}\right) A_{w,l} & \text{if } l \in \mathcal{U}^{\mathrm{c}} \\ 0 & \text{otherwise} \end{cases} \tag{47}$$

and $\mathcal{U}^{\mathrm{c}}$ stands for an complementary set of $\mathcal{U}$ in (46), i.e., $\mathcal{U}^{\mathrm{c}} \triangleq \{1, \ldots, N\} \setminus \mathcal{U}$. For the binary-input AWGN channel, the TSB on the conditional bit error probability admits the following final form (see [15]):

$$P_{\mathrm{b}|0}(\mathcal{C}'') \leq \int_{-\infty}^{\infty} \frac{dz_1}{\sqrt{2\pi}\sigma} e^{-\frac{z_1^2}{2\sigma^2}} \left\{ \begin{array}{l} \sum_{l:\frac{\delta_l}{2}\leq\alpha_l} \left\{ A_l{}'(\mathcal{C}'') \int_{\beta_l(z_1)}^{r_{z_1}} \frac{1}{\sqrt{2\pi}\sigma} e^{-\frac{z_2^2}{2\sigma^2}} \bar{\gamma}\left(\frac{N-2}{2}, \frac{r_{z_1}{}^2 - z_2^2}{2\sigma^2}\right) dz_2 \right\} \\ +1 - \bar{\gamma}\left(\frac{N-1}{2}, \frac{r_{z_1}{}^2}{2\sigma^2}\right) \end{array} \right\} \tag{48}$$

where the incomplete Gamma function $\bar{\gamma}$ is introduced in (20). As the simplest alternative to obtain an upper bound on the conditional bit error probability of the subcode $\mathcal{C}'$ given that the all-zero codeword is transmitted, one may use the union bound (UB) for the binary-input AWGN channel

$$\begin{aligned} P_{\mathrm{b}|0}(\mathcal{C}'') &\leq \sum_{w=1}^{NR} \left(\frac{w}{NR}\right) \sum_{l \in \mathcal{U}^{\mathrm{c}}} A_{w,l}\, Q\left(\sqrt{\frac{2lRE_{\mathrm{b}}}{N_0}}\right) \\ &= \sum_{l=1}^{N} A'_l(\mathcal{C}'')\, Q\left(\sqrt{\frac{2lRE_{\mathrm{b}}}{N_0}}\right) \end{aligned} \tag{49}$$

where $E_{\mathrm{b}}$ is the energy per information bit and $\frac{N_0}{2}$ is the two-sided spectral power density of the additive noise.

In order to tighten the upper bound (45), we obtain the bit-error version of the MSFB (see Eq. (25)), by following the steps of the proof of Theorem 1. In a similar manner to the transition from (7) to (40), we just need to replace the terms $A_l(\mathcal{C}')$ in (25) with $A'_l(\mathcal{C}')$ to get the conditional modified SFB (MSFB) on the bit error probability of $\mathcal{C}'$, given the all-zero codeword is transmitted. The resulting upper bound is expressed in the following theorem:

**Theorem 3 (Modified SFB on the Bit Error Probability).** Let $\mathcal{C}$ be a binary linear block code of length $N$ and rate $R$, and let $A_{w,l}$ be the number of codewords of $\mathcal{C}$ which are encoded by information bits whose Hamming weight is $w$ and their Hamming weight after encoding is $l$ (where $0 \leq w \leq NR$ and $0 \leq l \leq N$). Let the code $\mathcal{C}$ be partitioned into two subcodes, $\mathcal{C}'$ and $\mathcal{C}''$, where $\mathcal{C}'$ contains all codewords of $\mathcal{C}$ with Hamming weight $l \in \mathcal{U} \subseteq \{1, 2, , \ldots, N\}$ and the all-zero codeword, and $\mathcal{C}''$ contains the all-zero codeword and all the other codewords of $\mathcal{C}$ which are not in $\mathcal{C}'$. Assume that the communication takes place over an MBIOS channel. Then, the bit error probability of $\mathcal{C}$ under ML decoding is upper bounded by

$$P_{\mathrm{b}} \leq P_{\mathrm{b}|0}(\mathcal{C}') + P_{\mathrm{b}|0}(\mathcal{C}'')$$

where

$$P_{\mathrm{b}|0}(\mathcal{C}') \leq 2^{-N\left(E_0(\rho) - \rho(R + \frac{\log(\alpha_{\mathrm{b}}(\mathcal{C}'))}{N})\right)} \left[\sum_{l \in \mathcal{U}} \binom{N}{l} \left(\frac{A(\rho)}{A(\rho)+B(\rho)}\right)^{l} \left(\frac{B(\rho)}{A(\rho)+B(\rho)}\right)^{N-l}\right]^{\rho}, \quad 0 \leq \rho \leq 1 \tag{50}$$

$$\alpha_{\mathrm{b}}(\mathcal{C}') \triangleq \max_{l \in \mathcal{U}} \frac{A'_l}{2^{-N(1-R)}\binom{N}{l}}, \qquad A'_l \triangleq \sum_{w=1}^{NR} \left(\frac{w}{NR}\right) A_{w,l}$$

and the functions $A, B, E_0$ are introduced in (26), (27) and (9), respectively. An upper bound on the conditional bit error probability for the subcode $\mathcal{C}''$, $P_{\mathrm{b}|0}(\mathcal{C}'')$, can be either a union bound (49), the TSB (48) or any other improved bound.

*Discussion:* Note that $\alpha_{\mathrm{b}}(\mathcal{C}') \leq \alpha(\mathcal{C}')$, therefore the bound on the bit error probability in (50) is always smaller than the bound on the block error probability in (25), as one could expect.

In the derivation of the MSFB on the conditional block and bit error probabilities (see Eqs. (25) and (50), respectively), we obtain simplified expressions by taking out the maximum of $\left\{\frac{A_l(\mathcal{C}')}{B_l}\right\}$ and $\left\{\frac{A'_l(\mathcal{C}')}{B_l}\right\}$ from the corresponding summations in (29) and (43). This simplification was also done in [21] for the derivation of the SFB as a particular case of the DS2 bound. When considering the case of an upper bound on the block error probability, this simplification is reasonable because we consider the terms $\left\{\frac{A_l(\mathcal{C}')}{B_l}\right\}$ which vary slowly over a large range of the Hamming weights $l$ (see Fig. 2(a) when referring to ensembles of turbo-like codes whose average distance spectrum resembles the binomial distribution). However, by considering the terms $\left\{\frac{A'_l(\mathcal{C}')}{B_l}\right\}$ whose values change considerably with $l$ (see Fig. 2(b)), such simplification previously done for the block error analysis (i.e., taking out the maximal value of $\frac{A'_l(\mathcal{C}')}{B_l}$ from the summation) is expected to significantly reduce the tightness of the bound on the *bit error probability*. Thus, the modification which results in (50) does not seem in general to yield a good upper bound.[2] In order to get a tighter upper bound on the bit error probability we introduce the following theorem:

**Theorem 4 (Simplified DS2 Bound).** Let $\mathcal{C}$ be a binary linear block code of length $N$ and rate $R$, and let $A_{w,l}$ designate the number of codewords which are encoded by information bits whose Hamming weight is $w$ and their Hamming weight after encoding is $l$ (where $0 \leq w \leq NR$ and $0 \leq l \leq N$). Let the code $\mathcal{C}$ be partitioned into two subcodes, $\mathcal{C}'$ and $\mathcal{C}''$, where $\mathcal{C}'$ contains all the codewords in $\mathcal{C}$ with Hamming weight $l \in \mathcal{U} \subseteq \{1, 2, , \ldots, N\}$ and the all-zero codeword, and $\mathcal{C}''$ contains all the other codewords of $\mathcal{C}$ and the all-zero codeword. Let

$$A_l{}'(\mathcal{C}') \triangleq \begin{cases} \sum_{w=1}^{NR} \left(\frac{w}{NR}\right) A_{w,l} & \text{if } l \in \mathcal{U} \\ 0 & \text{otherwise} \end{cases} .$$

Assume that the communication takes place over an MBIOS channel. Then, under ML decoding, the bit error probability of $\mathcal{C}$, is upper bounded by

$$P_{\text{b}} \leq P_{\text{b}|0}(\mathcal{C}') + P_{\text{b}|0}(\mathcal{C}'')$$

where

$$P_{\text{b}|0}(\mathcal{C}') \;\leq\; 2^{-N\left(E_0(\rho)-\rho\left(R+\frac{\log \bar{\alpha}_\rho(\mathcal{C}')}{N}\right)\right)}, \qquad 0 \leq \rho \leq 1 \tag{51}$$

$$\bar{\alpha}_\rho(\mathcal{C}') \triangleq \sum_{l=0}^{N} \left\{ \frac{A'_l(\mathcal{C}')}{2^{-N(1-R)}\binom{N}{l}} \cdot \binom{N}{l} \left(\frac{A(\rho)}{A(\rho)+B(\rho)}\right)^l \left(\frac{B(\rho)}{A(\rho)+B(\rho)}\right)^{N-l} \right\}. \tag{52}$$

$A(\rho), B(\rho)$ and $E_0$ are defined in (26), (27) and (9), respectively. As before, an upper bound on the conditional bit error probability for the subcode $\mathcal{C}''$, $P_{\text{b}|0}(\mathcal{C}'')$, can be either a union bound or any other improved bound.

*Proof.* Starting from the first equality in (43), and using the definition for $A(\rho)$, $B(\rho)$ in

[2]Note that for an ensemble of fully random block codes, all the terms $\frac{A'_l}{B_l}$ are equal to $\frac{1}{2}$; hence, the simplification above does not reduce the tightness of the bound at all when considering this ensemble.

(26) and (27) we get

$$
\begin{aligned}
P_{\mathrm{b}|0} &\leq \left(\frac{A(\rho)+B(\rho)}{2}\right)^N 2^{N\rho} \left\{\sum_{l=0}^{N} A'_l(\mathcal{C}') \left(\frac{B(\rho)}{A(\rho)+B(\rho)}\right)^{N-l} \left(\frac{A(\rho)}{A(\rho)+B(\rho)}\right)^{l}\right\}^{\rho} \\
&= \left(\frac{A(\rho)+B(\rho)}{2}\right)^N 2^{NR\rho} \cdot 2^{N\rho(1-R)} \\
&\quad \cdot \left\{\sum_{l=0}^{N} A'_l(\mathcal{C}') \left(\frac{B(\rho)}{A(\rho)+B(\rho)}\right)^{N-l} \left(\frac{A(\rho)}{A(\rho)+B(\rho)}\right)^{l}\right\}^{\rho} \\
&= 2^{-N(E_0(\rho)-\rho R)} \left\{\sum_{l=0}^{N} \frac{A'_l(\mathcal{C}')}{B_l} \binom{N}{l} \left(\frac{B(\rho)}{A(\rho)+B(\rho)}\right)^{N-l} \left(\frac{A(\rho)}{A(\rho)+B(\rho)}\right)^{l}\right\}^{\rho}
\end{aligned}
\tag{53}
$$

where

$$
B_l \triangleq 2^{-N(1-R)} \binom{N}{l}, \qquad l = 0, \ldots, N
$$

designates the distance spectrum of fully random block codes of length $N$ and rate $R$. Using the definition for $\bar{\alpha}_\rho(\mathcal{C}')$ in (52) we get the upper bound (51). □

Evidently, the upper bound (51) is tighter than the bit-error version of the SFB in (45), because $\bar{\alpha}_\rho(\mathcal{C}')$ which is the expected value of $\frac{A'_l(\mathcal{C}')}{B_l}$ is not larger than $\alpha_{\mathrm{b}}(\mathcal{C}')$ which is the maximal value of $\frac{A'_l(\mathcal{C}')}{B_l}$. We note that the upper bound (51) is just the DS2 bound [8], with the un-normalized tilting measure (14). This tilting measure is optimal only for the ensemble of fully random block codes, and is sub-optimal for other codes. We refer to the upper bound (51) as the *simplified DS2*. From the discussion above, we conclude that the simplified DS2 bound (which is also valid as an upper bound on the conditional *block* error probability if we replace $A'_l(\mathcal{C}')$ in (53) by $A_l(\mathcal{C}')$) is advantageous over the MSFB when $A'_l$ (or $A_l$ for the case of block error probability) changes considerably over the Hamming weight range of interest. This is demonstrates for the block error probability of the ensemble of multiple turbo-Hamming codes where there is no noticeable improvement if we use the simplified DS2 to bound $P_{\mathrm{e}|0}(\mathcal{C}')$ instead of the MSFB, where for the case of bit-error probability we get tighter upper bound when using the simplified DS2 to upper bound $P_{\mathrm{b}|0}(\mathcal{C}')$ rather than the MSFB.

# 4 Expurgation

In this section we consider a possible expurgation of the distance spectrum which yields in general tighter upper bounds on the ML decoding error probability when transmission takes place over a binary-input AWGN (BIAWGN) channel. To this end, we rely on some properties of the Voronoi regions of binary linear block codes, as presented in [1, 2, 3].

Let $\mathcal{C}$ be a binary linear block code of length $N$ and rate $R$. Without any loss of generality, let us assume that the all-zero codeword, $\mathbf{c}_0$, was transmitted over the BIAWGN channel. For any received vector $\mathbf{y}$, an ML decoder checks whether it falls within the decision region of the all zero vector. This decision region (which is also called the Voronoi region of $\mathbf{c}_0$) is defined as the set $\mathcal{V}_0$ of vectors in $\mathbb{R}^N$ that are closest (in terms of Euclidian distance) to the all-zero codeword, i.e.,

$$\mathcal{V}_0 = \left\{\mathbf{x} \in \mathbb{R}^N : d(\mathbf{x}, \mathbf{c}_0) \leq d(\mathbf{x}, \mathbf{c}), \quad \forall\, \mathbf{c} \in \mathcal{C}\right\}. \tag{54}$$

Not all of the $2^{NR}$ inequalities in (54) are necessarily required to define the Voronoi region. The minimal set of codewords that determine the Voronoi region of $\mathbf{c}_0$, forms the set of Voronoi neighbors of $\mathbf{c}_0$ (to be designated by $\mathcal{N}_0$). So the region (54) can be defined by

$$\mathcal{V}_0 = \left\{\mathbf{x} \in \mathbb{R}^N : d(\mathbf{x}, \mathbf{c}_0) \leq d(\mathbf{x}, \mathbf{c}), \quad \forall\, \mathbf{c} \in \mathcal{N}_0\right\}. \tag{55}$$

It is clear that the block error probability of $\mathcal{C}$ is equal to the conditional block error probability of the expurgated subcode $\mathcal{C}^{\text{ex}}$, assuming the all-zero codeword is transmitted, where $\mathcal{C}^{\text{ex}}$ designates the subcode of $\mathcal{C}$ which contains the all-zero codeword and all its (Voronoi) neighbors. Hence, any upper bound that solely depends on the code distance spectrum of the code can be tightened by replacing the original distance spectrum with *the distance spectrum of the expurgated code.* It should be noted, however, that the argument above cannot be applied to the *bit* error probability. This stems from the fact that while the block error event is solely defined by the Voronoi region of the transmitted codeword, the bit error event also depends on the Hamming weight of the information bits of each decoded codeword; hence, the above expurgation cannot be applied to the analysis of the bit error probability. The distance spectrum of the Voronoi neighbors of an arbitrary codeword of some popular linear block codes (e.g., Hamming, BCH and Golay codes) is given in [1]. A simple way to find a subcode of $\mathcal{C}$ which contains the subcode $\mathcal{C}^{\text{ex}}$ is given in the following theorem from [2]:

**Theorem 5 (On the Voronoi Regions of Binary Linear Block Codes [2]).** For any binary linear block code $\mathcal{C}$ with rate $R$ and length $N$

$$\mathcal{N}_0 \supseteq \{\mathbf{c} \in \mathcal{C} : 1 \leq W_{\text{H}}(\mathbf{c}) \leq 2d_{\min} - 1\}$$

and

$$\mathcal{N}_0 \subseteq \{\mathbf{c} \in \mathcal{C} : 1 \leq W_{\text{H}}(\mathbf{c}) \leq N(1-R) + 1\}$$

where $d_{\min}$ is the minimal Hamming weight of the codewords in $\mathcal{C}$.

Note that according to the theorem above, one should expect the expurgation to have maximal impact on the tightness of an upper bound for high rate codes, where most of the codewords can be expurgated. We should also observe that the expurgated codewords have large distances from the all-zero codeword (all the expurgated codewords have a Hamming weight larger than $2d_{\min} - 1$). Thus, the improvement due to the expurgation process is especially substantial at low SNRs. One can use this theorem to achieve an immediate improvement of an arbitrary upper bound by expurgating all the codewords whose Hamming weight is greater than $N(1-R)+1$. We refer to this kind of expurgation as the *trivial* expurgation. The trivial expurgation, though very simple to apply, does not produce satisfactory results in many cases, because in many cases, the portion of the distance spectrum which corresponds to Hamming weights above $N(1-R)+1$ has a negligible effect on the overall bound. In [2], Agrell introduces a method (called *C rule*) in order to determine whether a codeword $\mathbf{c}$ is a zero-neighbor.

*C rule*: A codeword is a 0-neighbor if and only if it covers[3] no other nonzero codeword.

In [3] , Ashikmin and Barg used this rule to derive explicit formulas for the weight spectrums of zero-neighbors for various codes. This includes the families of Hamming codes and second-order Reed-Muller codes.

In order to upper bound the block error probability using the bounding technique introduced in this paper, we split the subcode $\mathcal{C}_{\text{ex}}$ into two subcodes, $\mathcal{C}'_{\text{ex}}$ and $\mathcal{C}''_{\text{ex}}$, where $\mathcal{C}'_{\text{ex}}$ contains all the codewords of $\mathcal{C}_{\text{ex}}$ with Hamming weight $l \in \mathcal{U} \subseteq \{1, 2, ..., N(1-R)+1\}$, and $\mathcal{C}''_{\text{ex}}$ contains the all-zero codeword and all the other codewords. The following upper bound holds:

$$P_{\text{e}}(\mathcal{C}) = P_{\text{e}|0}(\mathcal{C}_{\text{ex}}) \leq P_{\text{e}|0}(\mathcal{C}'_{\text{ex}}) + P_{\text{e}|0}(\mathcal{C}''_{\text{ex}}) \tag{56}$$

were $P_{\text{e}|0}(\mathcal{C}'_{\text{ex}})$ and $P_{\text{e}|0}(\mathcal{C}''_{\text{ex}})$ are the conditional block error probabilities of the subcodes $\mathcal{C}'_{\text{ex}}$ and $\mathcal{C}''_{\text{ex}}$, respectively, given that the all-zero codeword was transmitted. We can upper bound $P_{\text{e}|0}(\mathcal{C}''_{\text{ex}})$ by the union bound or the TSB, and we upper bound $P_{\text{e}|0}(\mathcal{C}'_{\text{ex}})$ by the MSFB (25). The partitioning of the subcode $\mathcal{C}_{\text{ex}}$ into two subcodes $\mathcal{C}'_{\text{ex}}$ and $\mathcal{C}''_{\text{ex}}$ is done following the adaptive algorithm introduced in Section 3.

# 5 Applications

This section demonstrates some numerical results of the improved upper bounds on the ML decoding error probability of linear block codes. We apply the bounds introduced in Sections 3 and 4 to various ensembles of parallel and serially concatenated codes. Throughout this section, it is assumed that the encoded bits are BPSK modulated, transmitted over an AWGN channel, and coherently detected. The effect of an expurgation of the distance spectrum on the tightness of some upper bounds on the decoding error probability is exemplified as well.

For the binary-input additive white Gaussian noise (BIAWGN) channel with BPSK modulation, the conditional probability density function (*pdf*) for a single letter input is:

$$\begin{aligned} p(y|0) &= \frac{1}{\sqrt{\pi N_0}} \exp\left\{-\left(y+\sqrt{E_{\text{s}}}\right)^2/N_0\right\}, \\ p(y|1) &= \frac{1}{\sqrt{\pi N_0}} \exp\left\{-\left(y-\sqrt{E_{\text{s}}}\right)^2/N_0\right\} \end{aligned} \tag{57}$$

where $E_{\text{s}}$ designates the energy of the symbol, and $\frac{N_0}{2}$ is the two-sided spectral power density of the channel. In order to calculate the SFB on $P_{\text{e}|0}(\mathcal{C}')$, we first calculate the terms $A(\rho)$ and $B(\rho)$, as defined in (26) and (27), respectively. Clearly, for a continuous-output channel,

[3]A binary codeword $\mathbf{c}_1$ is said to *cover* another codeword, $\mathbf{c}_2$, if $\mathbf{c}_2$ has zeros in all the positions where $\mathbf{c}_1$ has a zero.

the sums in (26) and (27) are replaced by integrals.

$$
\begin{aligned}
B(\rho) &= \int_{-\infty}^{\infty} p(y|0)^{\frac{2}{\rho+1}} \left[\frac{1}{2}p(y|0)^{\frac{1}{1+\rho}} + \frac{1}{2}p(y|1)^{\frac{1}{1+\rho}}\right]^{\rho-1} dy \\
&= \int_{-\infty}^{\infty} \left(\frac{1}{\sqrt{\pi N_0}}\right)^{\frac{2}{\rho+1}} e^{-\frac{2(y+\sqrt{E_s})^2}{N_0(1+\rho)}} \left(\frac{1}{\sqrt{\pi N_0}}\right)^{\frac{\rho-1}{\rho+1}} \left[\frac{1}{2}e^{-\frac{(y+\sqrt{E_s})^2}{N_0(1+\rho)}} + \frac{1}{2}e^{-\frac{(y-\sqrt{E_s})^2}{N_0(1+\rho)}}\right]^{\rho-1} dy \\
&= \exp\left(-\frac{E_s}{N_0}\right) \int_{-\infty}^{\infty} \frac{1}{\sqrt{\pi N_0}} e^{-\frac{y^2}{N_0}} \cdot e^{-\frac{4y\sqrt{E_s}}{N_0(\rho+1)}} \left[\frac{1}{2}e^{\frac{2y\sqrt{E_s}}{N_0(1+\rho)}} + \frac{1}{2}e^{-\frac{2y\sqrt{E_s}}{N_0(1+\rho)}}\right]^{\rho-1} dy \\
&= \exp\left(-\frac{E_s}{N_0}\right) \mathbb{E}\left[e^{-\frac{2X\sqrt{2E_s/N_0}}{\rho+1}} \cosh^{\rho-1}\!\left(\frac{\sqrt{2E_s/N_0}X}{1+\rho}\right)\right] \qquad (58)
\end{aligned}
$$

where $\mathbb{E}$ denotes the statistical expectation, and $X \sim \mathrm{N}(0,1)$. We also obtain that

$$
\begin{aligned}
A(\rho) &= \int_{-\infty}^{\infty} [p(y|0)p(y|1)]^{\frac{1}{1+\rho}} \left[\frac{1}{2}p(y|0)^{\frac{1}{1+\rho}} + \frac{1}{2}p(y|1)^{\frac{1}{1+\rho}}\right]^{\rho-1} dy \\
&= \exp\left(-\frac{E_s}{N_0}\right) \mathbb{E}\left[\cosh^{\rho-1}\left(\frac{\sqrt{2E_s/N_0}X}{1+\rho}\right)\right] \qquad (59)
\end{aligned}
$$

and

$$
A(\rho) + B(\rho) = 2\exp\left(-\frac{E_s}{N_0}\right) \mathbb{E}\left[\cosh^{1+\rho}\left(\frac{\sqrt{2E_s/N_0}X}{1+\rho}\right)\right] \qquad (60)
$$

Plugging (58) – (60) into (25), and (50) and minimizing over the interval $0 \le \rho \le 1$ will give us the desired bounds for $P_{\mathrm{e}|0}(\mathcal{C}')$ and $P_{\mathrm{b}|0}(\mathcal{C}')$, respectively.

## 5.1 Ensemble of Serially Concatenated Codes

The scheme in Fig. 3 depicts the encoder of an ensemble of serially concatenated codes where the outer code is a (127, 99, 29) Reed-Solomon (RS) code, and the inner code is chosen uniformly at random from the ensemble of (8, 7) binary linear block codes. Thus, the inner code extends every symbol of 7 bits from the Galois field $\mathrm{GF}(2^7)$ to a sequence of 8 bits. The decoding is assumed to be performed in two stages: the inner (8, 7) binary linear block code is soft-decision ML decoded, and then a hard decision ML decoding is used for the outer (129, 99, 29) RS code. Due to the hard-decision ML decoding of the (127, 99, 29) RS code, its decoder can correct up to $t = \lfloor \frac{d_{\min}-1}{2} \rfloor = 14$ erroneous symbols. Hence, an upper bound on the average block error probability of the considered serially concatenated ensemble is given by

$$
P_e \le \sum_{i=t+1}^{127} \binom{127}{i} p_s^i (1-p_s)^{127-i} \qquad (61)
$$

where $p_s$ is the average symbol error probability of the inner code under soft-decision ML decoding. The symbol error probability $p_s$ of the inner code is either upper bounded by the ubiquitous union bound or the TSB, and this upper bound is substituted in the RHS of (61). Since the rate of the inner code is rather high (it is equal to $\frac{7}{8}$ bits per channel use), an expurgation of the distance spectrum seems to be attractive in order to tighten the upper bound on the overall performance of the concatenated ensemble. Ashikmin and Barg [3] show that the average expurgated distance spectrum of the ensemble of random linear block codes of length $N$ and dimension $K$ is given by

$$E[A_l] = \begin{cases} \binom{N}{l} 2^{-(N-K)} \prod_{i=0}^{l-2} \left(1 - 2^{-(N-K-i)}\right) & l = 0, 1, \ldots, N-K+1 \\ 0 & \text{otherwise.} \end{cases} \quad (62)$$

We rely on the expurgated distance spectrum in (62) in order to get a tighter version of the union bound or the TSB on the symbol error probability $p_{\rm s}$ of the inner code (where $N = 8$ and $K = 7$). The expurgated union bound in Fig. 4 provides a gain of 0.1 dB over the union bound or TSB at block error probability of $10^{-4}$, and the improvement in the tightness of the bound due to the distance spectrum expurgation is especially prominent at low values of SNR. Clearly, we take 1 as the trivial bound on $p_{\rm s}$ (as otherwise, for low values of SNR, the union bound on $p_{\rm s}$ may exceed 1, which gives in turn a useless upper bound on the decoding error probability of the ensemble).

## 5.2 Turbo-Hamming Codes

Let us consider an ensemble of uniformly interleaved parallel concatenated turbo-Hamming codes. The encoder consists of two identical $(2^m - 1, 2^m - m - 1)$ Hamming codes as component codes, and a uniform interleaver operating on the $2^m - m - 1$ information bits. The comparison here refers to the case where $m = 10$, so the two component codes are (1023, 1013) Hamming codes, and the overall rate of the ensemble is $R = \frac{2^m - m - 1}{2^m + m - 1} = 0.9806$ bits per channel use. The value of the energy per bit to one-sided spectral noise density $(\frac{E_{\rm b}}{N_0})$ which corresponds to this coding tare is 5.34 dB, assuming that communication takes place over a binary-input AWGN channel. In order to obtain performance bounds for the ensemble of uniformly interleaved turbo-Hamming codes, we rely on an algorithm for the calculation of the average input-output weight enumerator function (IOWEF) of this ensemble, as provided in [17, Section 5.2]. As noted in [17], the average distance spectrum of this ensemble is very close to the binomial distribution for a rather large range of Hamming weights (see Fig. 2(a)). Hence, one can expect that the upper bound introduced in Theorem 1 provides a tight bounding technique on the average block error probability of this ensemble. For this coding scheme, we note that regarding $P_{\rm e}$, there is no substantial improvement in the tightness of the overall upper bound if we upper bound $P_{{\rm e}|0}(\mathcal{C}'')$ by the TSB instead of the simple union bound (see Fig. 6). Among the bounds introduced in Section 3, the upper bound which combines the TSB and the MSFB is the tightest bound, especially for the low SNR range (see Fig. 6); referring to the bound in Theorem 1, the partitioning of codes in the considered ensemble relies on Algorithm 1 (see Section 3). In Fig. 7, we provide a comparison between various upper bound on the *bit* error probability of this turbo-like ensemble. The tightest bound for the bit error analysis is the one provided in Theorem 4, combining the simplified DS2 bound with the union bound. It is shown in Fig. 7 that the simplified DS2 provides gains of 0.16 dB and 0.05 dB over the MSFB at bit error probabilities of $10^{-1}$ and $10^{-2}$, respectively. The simplified DS2 also provides gain of 0.08 dB over the TSB at bit error probability of $10^{-1}$. Unfortunately, a trivial expurgation of the average distance spectrum of uniformly interleaved turbo codes with two identical $(2^m - 1, 2^m - m - 1)$ Hamming codes as components (i.e., by nullifying the average distance spectrum at Hamming weights above $2m+1$) has no impact on tightening the performance bounds of this ensemble.

## 5.3 Multiple Turbo-Hamming Codes

Multiple turbo codes are known to yield better performance, and hence, it is interesting to apply the new bounding techniques in Section 3 to these ensembles. The encoder of a multiple turbo-Hamming code is depicted in Fig. 8.

Consider the ensemble of uniformly and independently interleaved multiple-turbo codes, where the components codes are identical systematic binary linear block codes of length $N$. Let $S_{w,h_i}$ denote the number of codewords of the $i^{\text{th}}$ component code with weight of the systematic bits equal to $w$ and the weight of the parity bits equal to $h_i$. The average number of codewords of the ensemble of multiple-turbo codes, with systematic-bits weight of $w$ and overall weight $l$ is given by

$$A_{w,l} = \sum_{\substack{h_1, h_2, h_3 \quad \text{s.t.} \\ w+h_1+h_2+h_3=l}} \frac{S_{w,h_1} S_{w,h_2} S_{w,h_3}}{\binom{N}{w}^2}. \tag{63}$$

From (63) and the algorithm to calculate the input-output weight enumerators of Hamming codes (see [17, Appendix A]), it is possible to verify that the average distance spectrum of the ensemble of multiple turbo-Hamming codes with two independent uniform interleavers is very close to the binomial distribution for a relatively large range of Hamming weights (similarly to the plot in Fig. 2(a)). Hence, the improved bounds provided in Section 3 are expected to yield good upper bounds on the decoding error probability. The comparison here refers to the case of $m = 10$, so the three component codes are (1023, 1013) Hamming codes. The overall rate of the ensemble is $\frac{2^m-m-1}{2^m+2m-1} = 0.9712$ bits per channel use, and the channel capacity for this coding rate corresponds to $\frac{E_{\text{b}}}{N_0} = 5$ dB. All the improved bounds that are evaluated here, incorporate the union bound as an upper bound on $P_{\text{e}}(\mathcal{C}'')$ (or $P_{\text{b}}(\mathcal{C}'')$ for bit error probabilities). The numerical results of various upper bounds are shown in Fig. 9 for the block and bit error probabilities. As expected, the improvements that were obtained by the improved bounds (Theorems 1–4) are more pronounced here than for the ensemble of turbo-Hamming code. For example, at bit error rate of $10^{-1}$, the simplified DS2 bound yields a gain of 0.12 dB over the TSB. A modest improvement of 0.05 dB was obtained at bit error rate of $10^{-2}$.

## 5.4 Random Turbo-Block Codes with Systematic Binary Linear Block Codes as Components

Finally, we evaluate improved upper bound for the ensemble of uniformly interleaved parallel concatenated (turbo) codes, having two identical component codes chosen uniformly at random and independently from the ensemble of systematic binary linear block codes. We assume that the parameters of the overall code are $(N, K)$, so the parameters of its component codes are $(\frac{N+K}{2}, K)$. In addition, the length of the uniform interleaver is $K$.

According to the analysis in [22], the input-output weight enumeration of the considered ensemble is given by

$$\begin{aligned} S(W,Z) &= \sum_{w,j} S_{w,j} W^w Z^j \\ &= 1 + \sum_{w=1}^{K} \left\{ W^w \left[ 2^{-(N-K)} \left( \binom{K}{w} - 1 \right) \sum_{j=0}^{N-K} \binom{N-K}{j} Z^j + 2^{-\frac{N-K}{2}} \sum_{j=0}^{\frac{N-K}{2}} \binom{\frac{N-K}{2}}{j} Z^{2j} \right] \right\} \end{aligned}$$

where $S_{w,j}$ denotes the number of codewords whose information sub-words have Hamming weight of $w$ and the parity sub-word has Hamming weight $j$. We apply the improved bounds introduced in Section 3 to this ensemble where the parameters are set to $(N, K) = (1144, 1000)$ (hence, the rate of the parallel concatenated ensemble is $R = 0.8741$ bits per channel use). The plots of various upper bounds on the block and bit error probabilities are shown in Fig. 10. The improved bounds yield the best reported upper bound on the block and bit error probabilities. For the block error probability, the upper bound which combines the MSFB with the union bound is the tightest bound; it achieve a gain of 0.1 dB over the TSB, referring to a block error probability of $10^{-4}$. A similar gain of 0.11 dB is obtained for the bit error probability, referring to a BER of $10^{-4}$, referring to the bound which combined the union bound with the simplified DS2 bound (see Theorem 4).

# 6 Conclusions

We derive in this paper tightened versions of the Shulman and Feder bound. The new bounds apply to the bit and block error probabilities of binary linear block codes under ML decoding. The effectiveness of these bounds is exemplified for various ensembles of turbo-like codes over the AWGN channel. An expurgation of the distance spectrum of binary linear block codes further tightens in some cases the resulting upper bounds.

# Figures

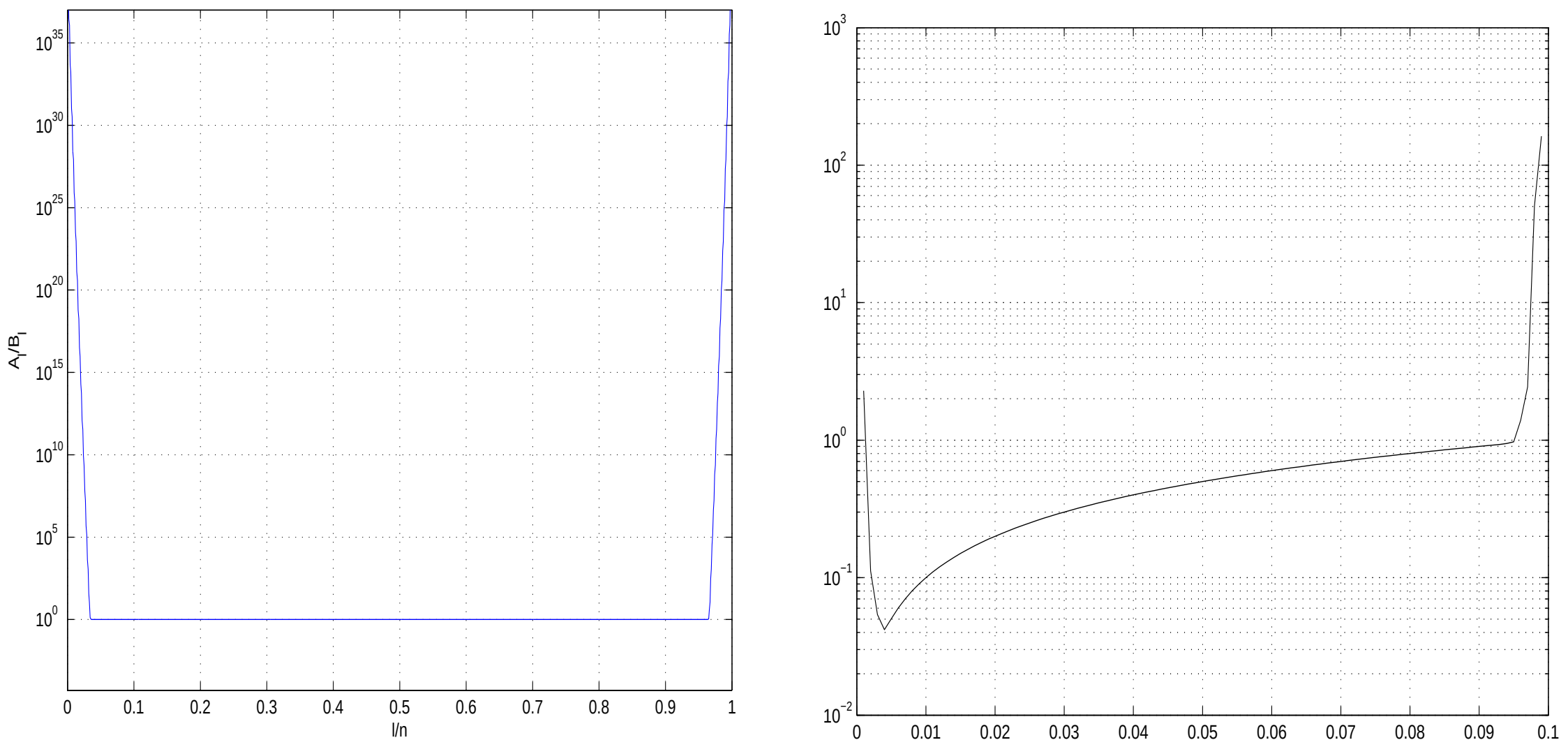

(a) (b)

Figure 2: Plots of $\frac{A_l}{B_l}$ and $\frac{A'_l}{B_l}$ as a function of the normalized Hamming weight $\left(\frac{l}{N}\right)$, on a logarithmic scale. The plots refer to ensembles of random turbo-block codes with two identical systematic binary linear block codes as components; (a) A plot of $\frac{A_l}{B_l}$ with $N = 1000$ and $R = 0.72$ bits/Symbol, referring to the analysis of the block error probability, (b) A plot of $\frac{A'_l}{B_l}$ with $N = 100$ and $R = 0.72$ bits/Symbol, referring to the analysis of the bit error probability.

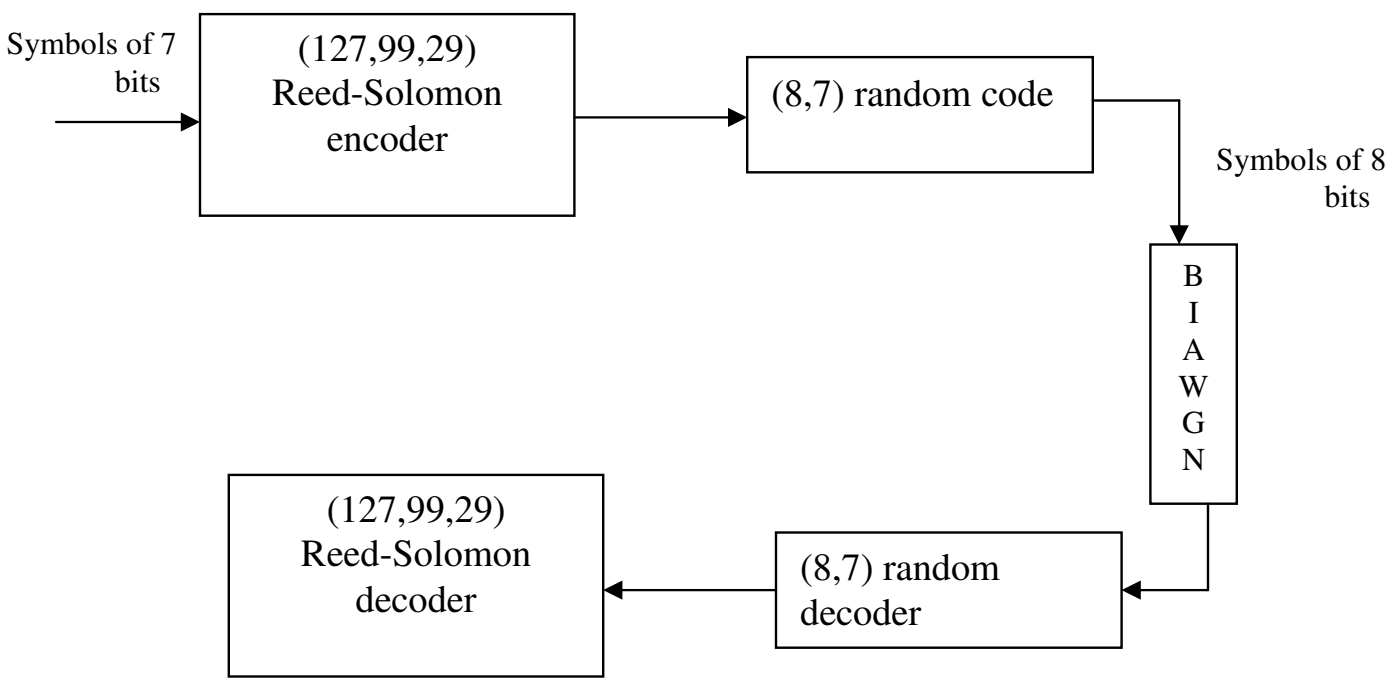

Figure 3: A scheme for an ensemble of serially concatenated codes where the outer code is a (127, 99, 29) Reed-Solomon (RS) code, and the inner code is chosen uniformly at random from the ensemble of (8,7) binary linear block codes.

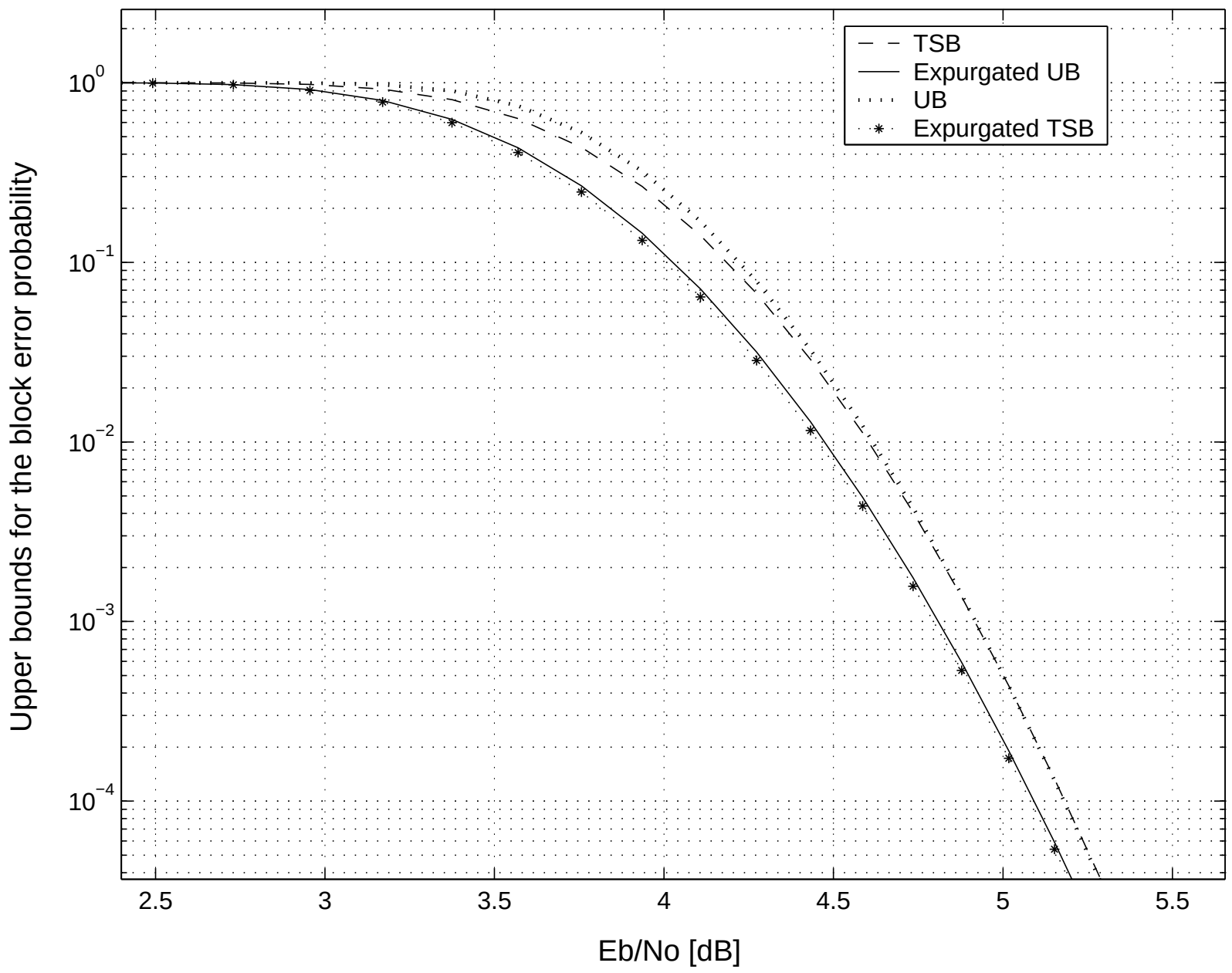

Figure 4: Various upper bounds on the block error probability of the ensemble of serially concatenated codes depicted in Fig. 3. The compared bounds are the tangential-sphere bound (TSB) and the union bound with and without expurgation of the distance spectrum; this expurgation refers to the ensemble of inner codes, chosen uniformly at random from the ensemble of (8,7) binary linear block codes.

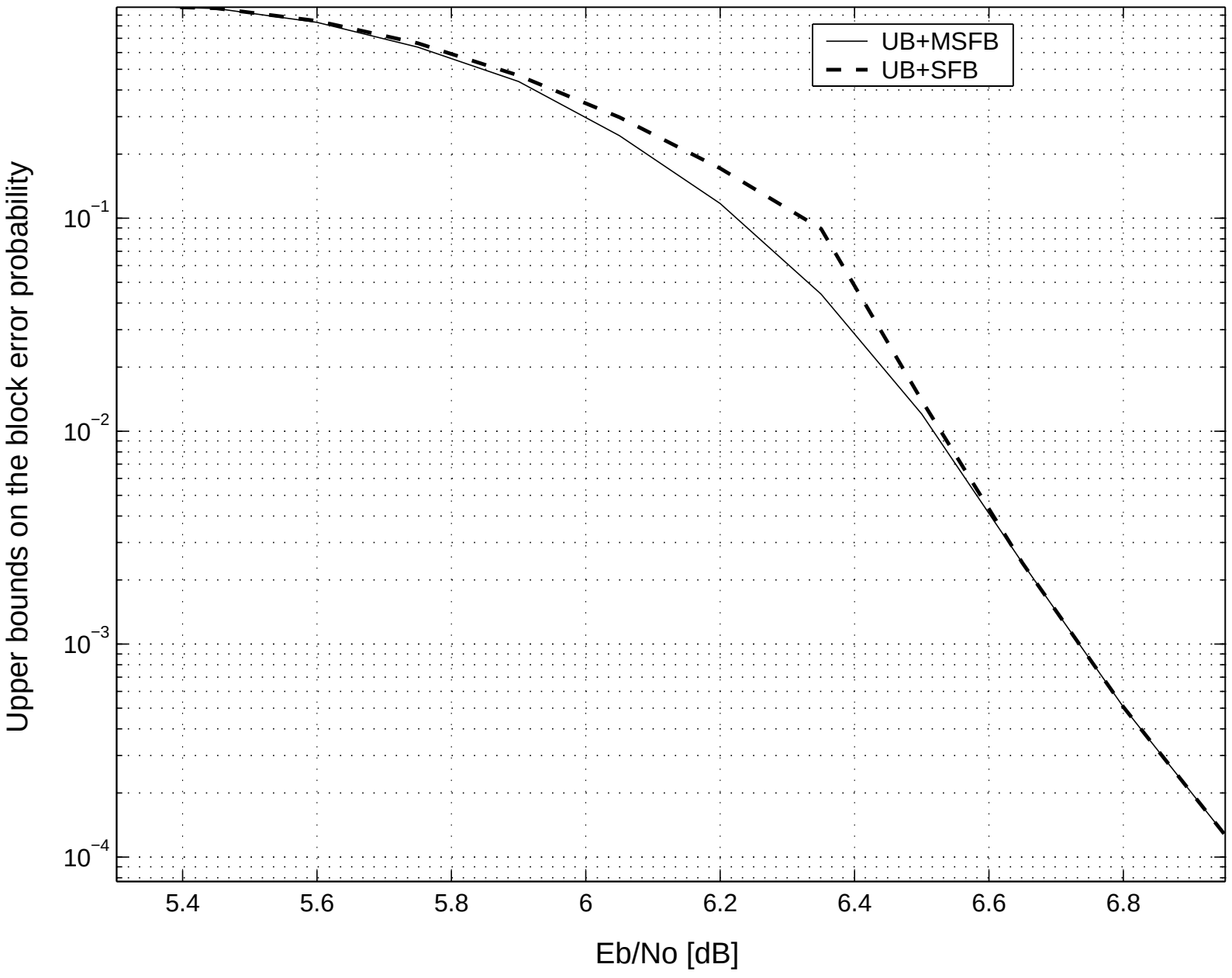

Figure 5: A comparison between the upper bound which combines the UB with the SFB bound in it original form (Eq. (7)) and the upper bound which combines the UB with the MSFB bound in (25). The comparison refers to the ensemble of uniformly interleaved turbo-Hamming codes where the two component codes are (1023, 1013) Hamming codes. The overall rate of the code is 0.973 bits per channel use.

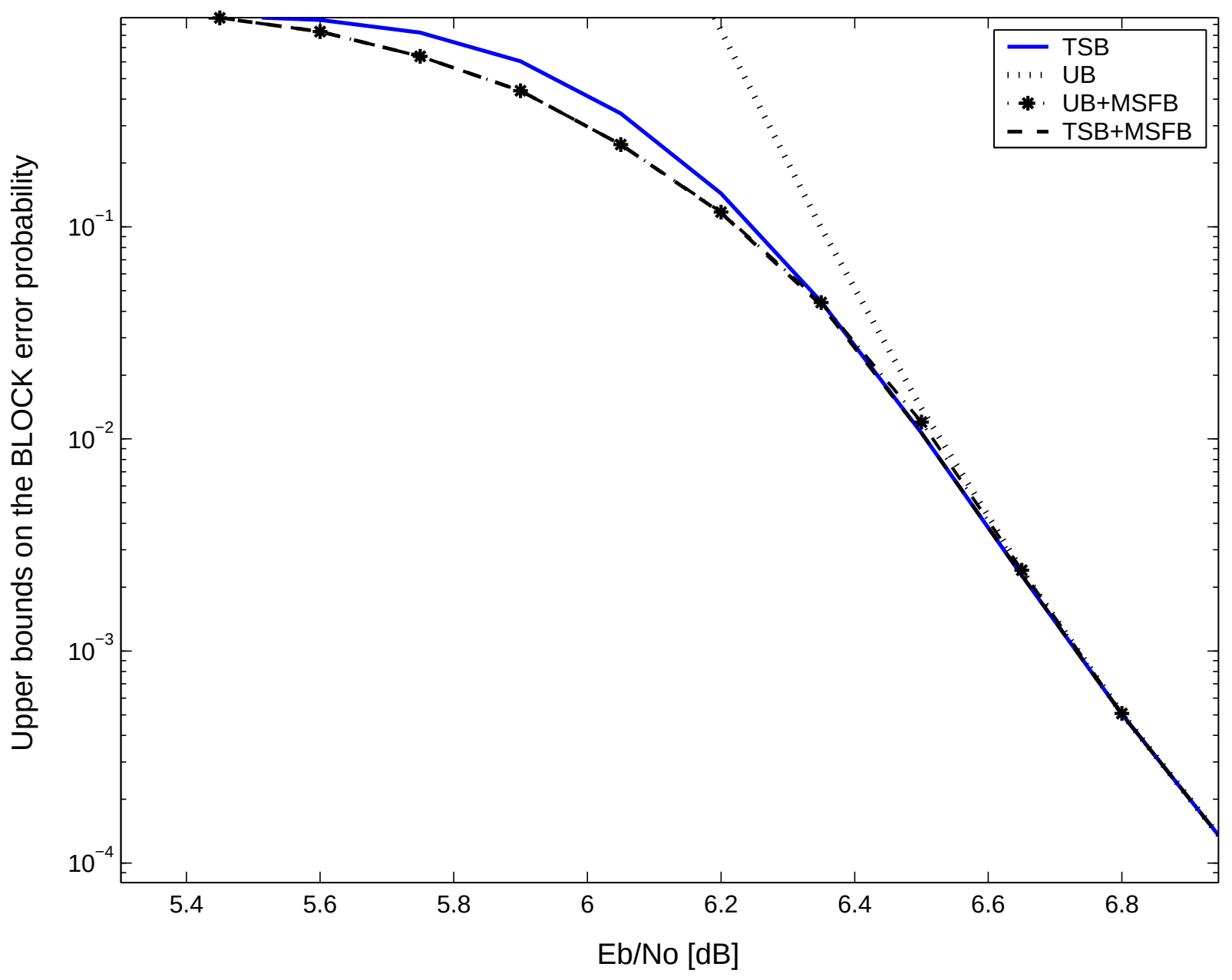

Figure 6: Comparison between various upper bounds on the ML decoding block error probability where the comparison refers to the ensemble of uniformly interleaved turbo-Hamming codes whose two component codes are (1023, 1013) Hamming codes. The compared bounds are the union bound (UB), the tangential-sphere bound (TSB), and two instances of the improved upper bound from Theorem 1: the UB+MSFB combines the MSFB with the union bound, and the TSB+MSFB is the upper bound which combines the MSFB with the tangential-sphere bound.

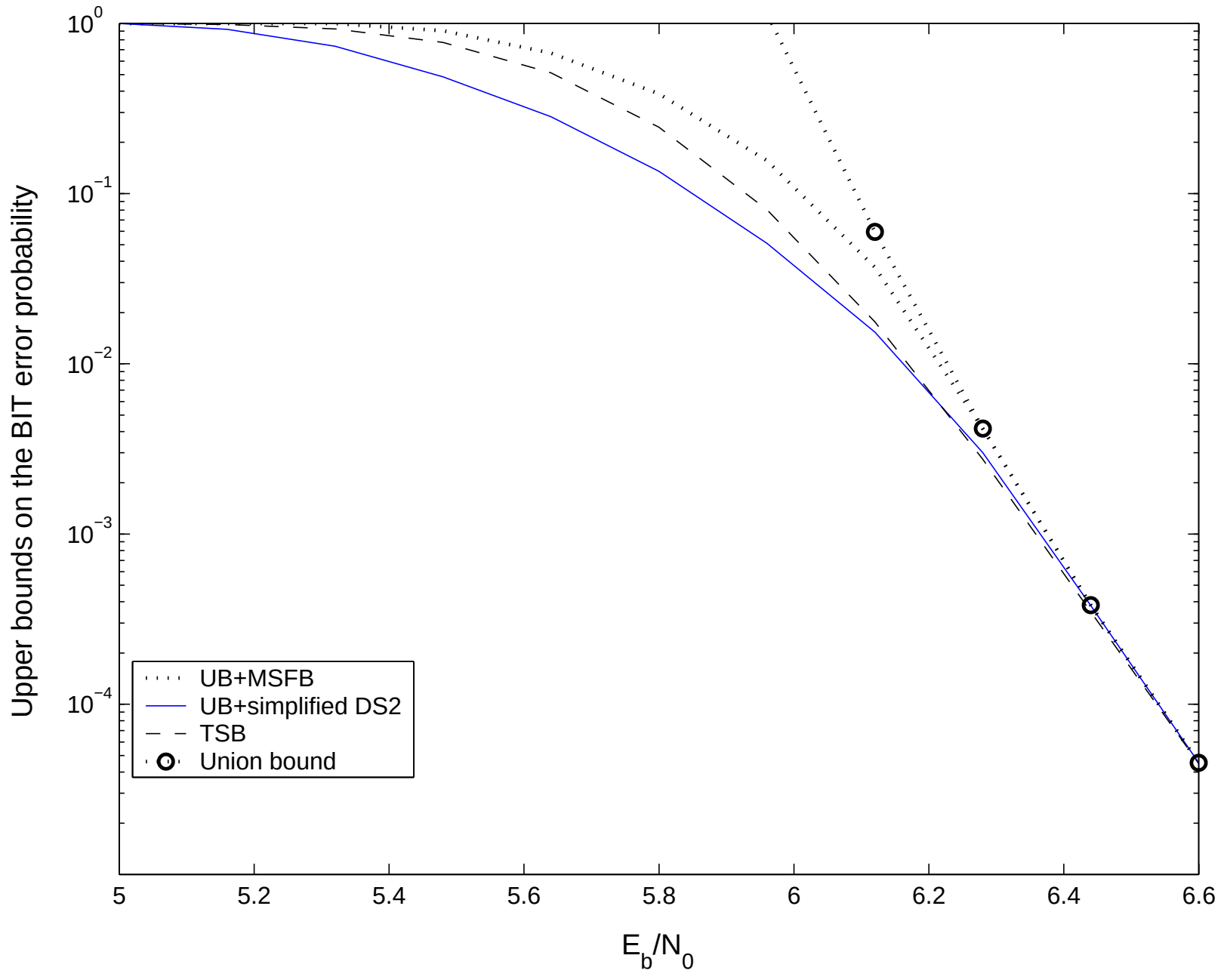

Figure 7: Comparison between various upper bounds on the ML decoding bit error probability of the ensemble of (1033,1013) uniformly interleaved turbo-Hamming code. The compared bounds are the union bound (UB), the tangential-sphere bound (TSB), the upper bound from Theorem 3 which combines the union bound with the MSFB (UB+MSFB), and the upper bound from Theorem 4 which combines the union bound with the simplified DS2 bound (UB+simplified DS2).

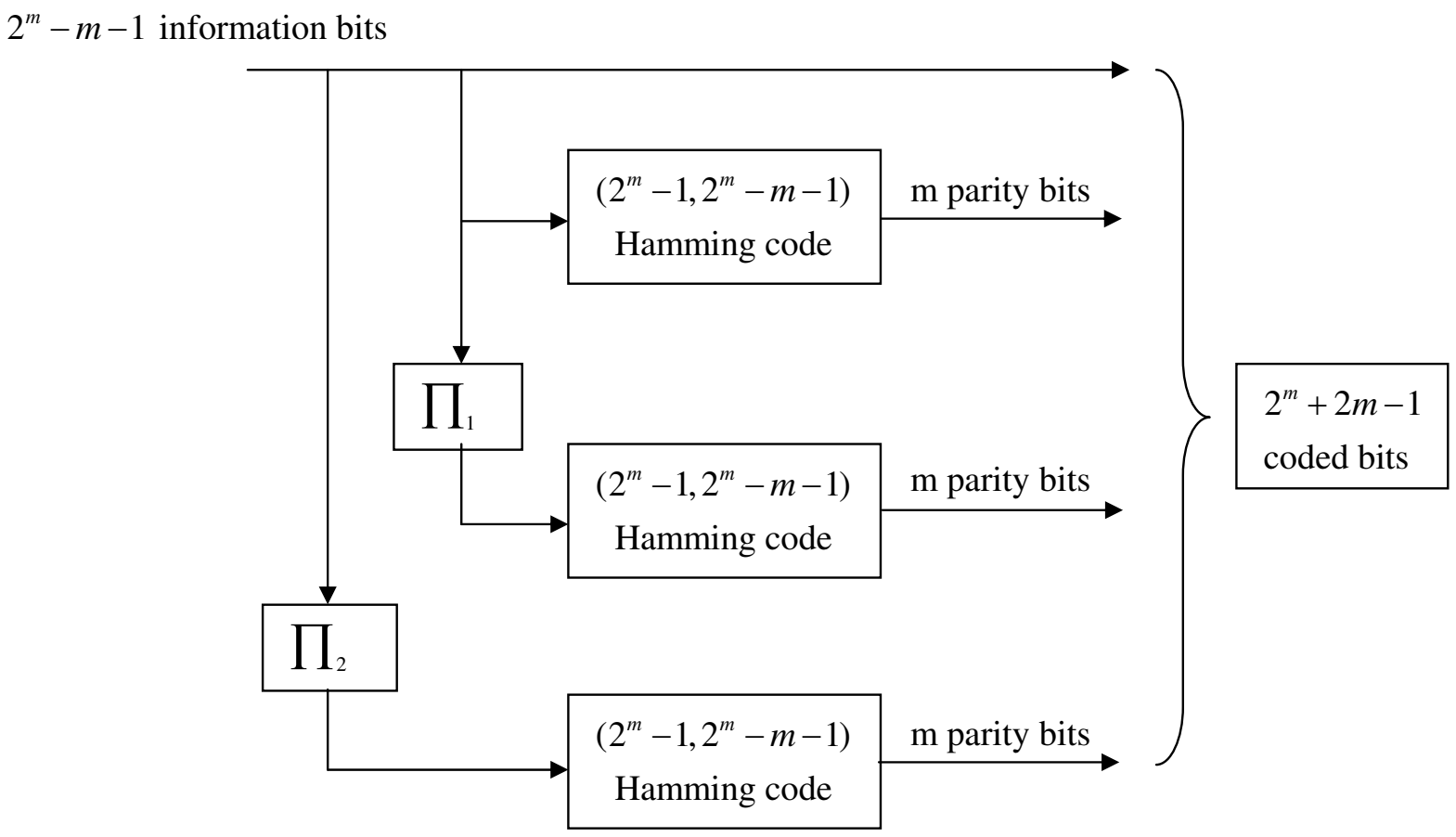

Figure 8: A multiple turbo-Hamming encoder. The encoder consists of parallel concatenated Hamming codes with two uniform, statistically independent interleavers. The code length is $2^m + 2m - 1$ and the code rate is $R = \frac{2^m - m - 1}{2^m + 2m - 1}$ bits per channel use.

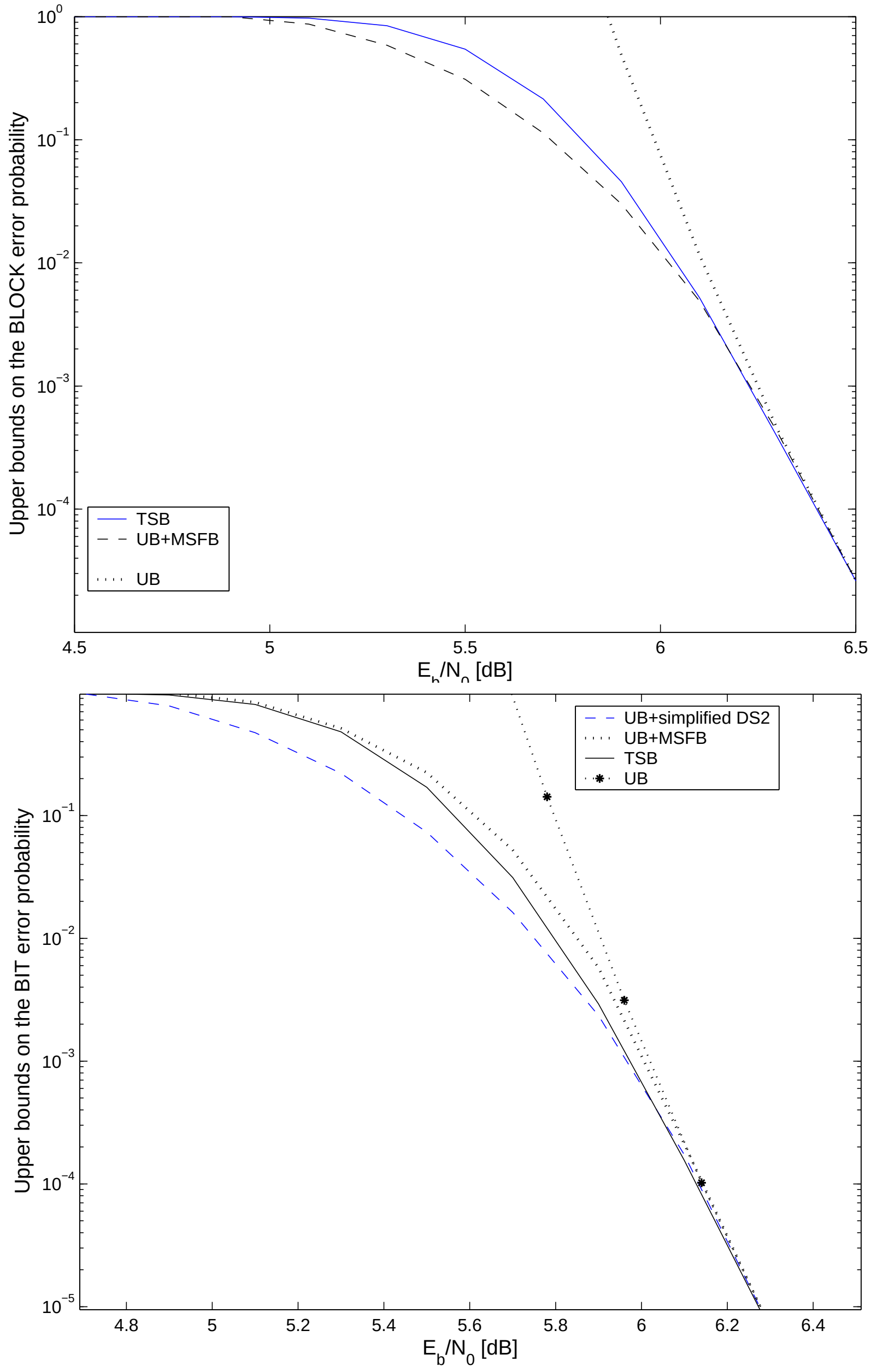

Figure 9: Comparison between various upper bounds on the ML decoding error probability, referring to the ensemble of uniformly interleaved multiple turbo-Hamming codes where the three component codes are (1023, 1013) Hamming codes (see Fig. 8). The upper plot refers to upper bounds on the block error probability, and the compared bounds are the union bound (UB), the tangential-sphere bound (TSB), and the upper bound of Theorem 1 which combines the union bound with the MSFB (UB+modified SFB). The lower plot refers to upper bounds on the bit error probability, and the compared bounds are the union bound (UB), the tangential-sphere bound (TSB), the upper bound of Theorem 3 which combines the union bound with the MSFB, and the upper bound of Theorem 4 which combines the union bound with the simplified DS2 bound (UB+simplified DS2).

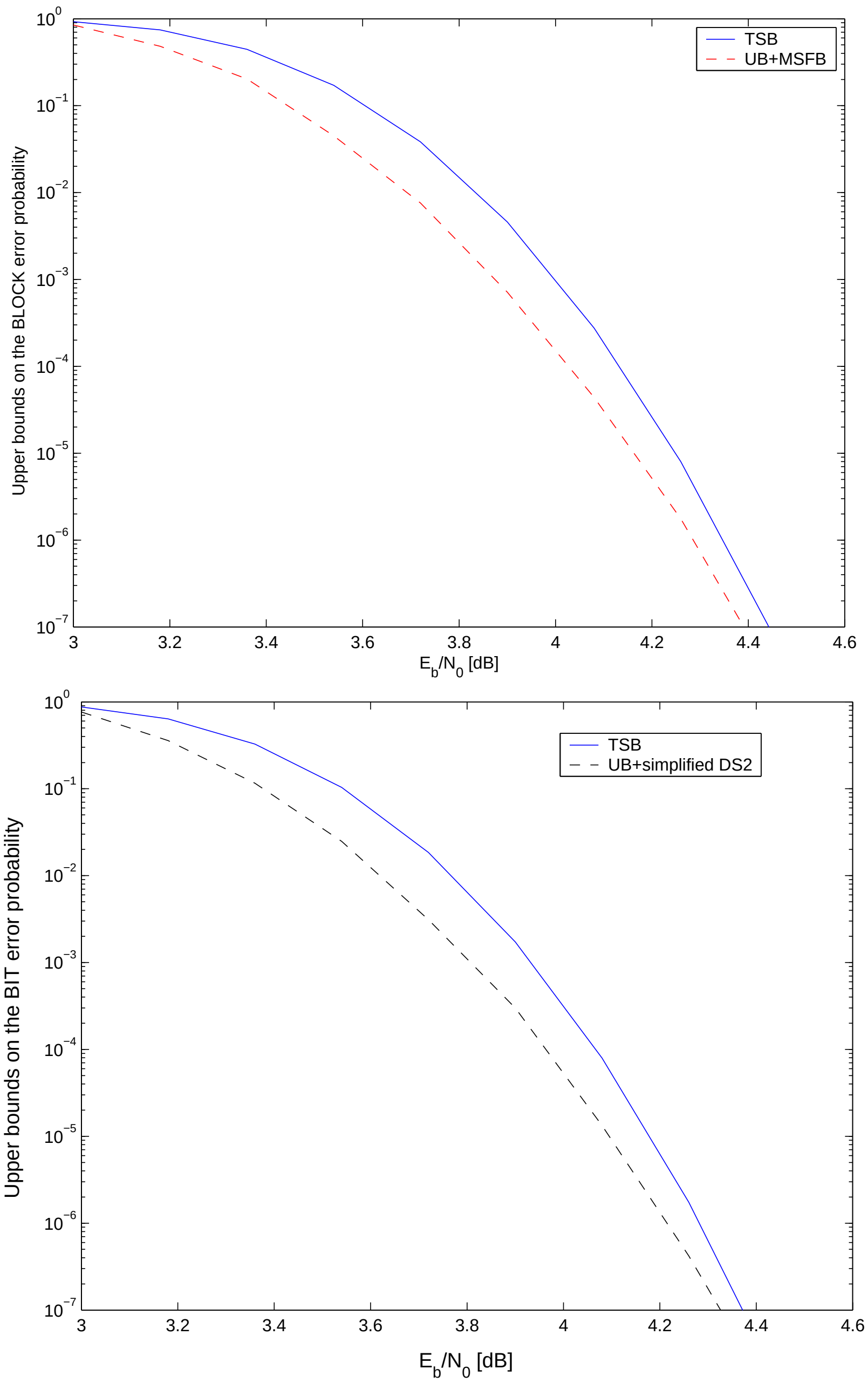

Figure 10: Comparison between upper bounds on the block and bit error probabilities for an ensemble of uniformly interleaved turbo codes whose two component codes are chosen uniformly at random from the ensemble of (1072, 1000) binary systematic linear block codes; its overall code rate is 0.8741 bits per channel use. The compared bounds under ML decoding are the tangential-sphere bound (TSB), and the bounds in Theorems 1 and 4. The upper and lower plots provide upper bounds on the block and bit error probabilities, respectively.